\documentclass[12pt, a4paper, parskip, DIV=15]{scrartcl}
\usepackage[english]{babel}
\usepackage[utf8]{inputenc}
\usepackage[T1]{fontenc}  \usepackage{textcomp, booktabs, multirow}
\usepackage[hidelinks]{hyperref}
\usepackage{graphicx, grffile}  \usepackage[separate-uncertainty=true]{siunitx}
\usepackage{amsmath, xspace}

\usepackage{lmodern}
\usepackage{doi, authblk}
\usepackage[round]{natbib}
\usepackage[format=plain, indention=0.5cm, font=small]{caption}
\usepackage{sidecap}

\usepackage{xcolor}

\catcode`\^^M=10

\newcommand{\ind}[1]{_\text{#1}}
\renewcommand{\vec}{\boldsymbol}
\newcommand{\parent}[1]{\ensuremath{\left(#1\right)}\xspace}
\newcommand{\fof}{\!\parent}
\newcommand{\sca}{\ind{sc}}
\newcommand{\intr}{\ind{i}}

\newcommand{\md}{\ind{mod}}
\newcommand{\obs}{\ind{obs}}
\newcommand{\omM}{\omega\hspace{-0.1em}M}
\newcommand{\omMf}{\omM\fof f}

\newcommand{\Ml}{\ensuremath{M\ind L}\xspace}
\newcommand{\Mw}{\ensuremath{M\ind w}\xspace}
\newcommand{\fc}{\ensuremath{f\ind c}\xspace}

\hypersetup{pdfauthor={Tom Eulenfeld et al.}, pdftitle={Earthquake source parameters, attenuation, and site effects from waveform envelopes}}

\title{\LARGE Induced earthquake source parameters, attenuation, and site effects from waveform envelopes in the Fennoscandian Shield}
\author[1,*]{Tom Eulenfeld}
\author[2]{Gregor Hillers}
\author[2]{Tommi A. T. Vuorinen}
\author[1]{Ulrich Wegler}
\affil[1]{Institute of Geosciences, Friedrich Schiller University Jena, 07749 Jena, Germany}
\affil[2]{Institute of Seismology, University of Helsinki, 00014 Helsinki, Finland}
\affil[*]{contact: tom.eulenfeld@uni-jena.de}
\date{{\vspace{-0.5cm}\small \today}}
\begin{document}
\maketitle

\vspace{-1.2cm}
\begin{center}
\color{gray} \footnotesize 
An edited version of this article was published by\\\emph{Journal of Geophysical Research: Solid Earth}, doi:\href{http://doi.org/10.1029/2022JB025162}{10.1029/2022JB025162}.
\end{center}
\vspace{-0.5cm}

\begin{abstract}\noindent
We analyze envelopes of 233 and 22 {\Ml}0.0 to {\Ml}1.8 earthquakes induced by two geothermal stimulations in the Helsinki, Finland, metropolitan area. We separate source spectra and site terms and determine intrinsic attenuation and the scattering strength of shear waves in the \SI{3}{Hz} to \SI{200}{Hz} frequency range using radiative transfer based synthetic envelopes. Displacement spectra yield scaling relations with a general deviation from self-similarity, with a stronger albeit more controversial signal from the weaker 2020 stimulation. The 2020 earthquakes also tend to have a smaller local magnitude compared to 2018 earthquakes with the same moment magnitude. We discuss these connections in the context of fluid effects on rupture speed or medium properties. Site terms demonstrate that the spectral amplification relative to two reference borehole sites is not neutral at the other sensors; largest variations are observed at surface stations at frequencies larger than \SI{30}{Hz}. Intrinsic attenuation is exceptionally low with $Q\intr^{-1}$ values down to \num{2.4e-5} at \SI{20}{Hz}, which allows  the observation of a diffuse reflection at the ${\sim}\SI{50}{km}$ deep Moho. Scattering strength is in the range of globally observed data with $Q\sca^{-1}$ between \num{e-3} and \num{e-4}.
The application of the employed Qopen analysis program to the 2020 data in a retrospective monitoring mode demonstrates its versatility as a seismicity processing tool. The diverse results have implications for scaling relations, hazard assessment and ground motion modeling, and imaging and monitoring using ballistic and scattered wavefields in the crystalline Fennoscandian Shield environment.

\newpage
\paragraph{Plain language summary}
We analyze seismograms from earthquakes that were induced during two geothermal stimulation experiments in the Helsinki, Finland, metropolitan area, in 2018 and 2020. We process long signals including later parts of the seismograms to solve the persistent problem of separating the effects of the earthquake source process, of the bedrock, and of the ground immediately below a seismic sensor on the observed data. The high data quality allows us to measure systematic differences in some fundamental earthquake source parameters between events induced during the two stimulations. We attribute this to the effect of the fluids that were pumped into the \SI{6}{km} deep rock formations. These observations are important since natural earthquakes and earthquakes induced by such underground engineering activities are governed by the same physical mechanisms. We also find that the bedrock in southern Finland is characterized by some of the lowest seismic attenuation values that have so far been measured in different tectonic environments. Last, the so-called site effects at the instrument locations show a diverse amplification pattern in a wide frequency range, which is important for the assessment of shaking scenarios in the area.

\newline
\newline
Key points:
\begin{itemize}
\item We find lower stress drop values for events induced by the 2020 compared to the 2018 stimulation and a deviation from self-similar scaling
\item The observation of a diffuse reflection at the \SI{50}{km} deep Moho highlights the low intrinsic attenuation in the Fennoscandian Shield
\item Site effect terms between \SI{3}{Hz} and \SI{200}{Hz} show diverse frequency and site dependent patterns with high-frequency amplification
\end{itemize}
\vspace{1ex}
\end{abstract}

\section{Introduction}
\label{sec:intro}
Induced and natural earthquakes display a wide range of complex behavior but are understood to be governed by the same physics.
Estimates of earthquake source properties are potentially biased by incomplete or uneven observations of natural seismic activity \citep{Ruhl2017}.
The stimulation of an enhanced geothermal system for energy production constitutes an in-situ laboratory, and the controllable aspects of an injection experiment including source region and timing facilitate the tuning of seismic monitoring networks for consistent analysis of the induced seismicity.
Observations of small induced event sequences using modern dense networks can thus have general implications for earthquake science and for the structural properties of the stimulation environment.
\par
\begin{figure}
\centering
\includegraphics[width=0.68\textwidth]{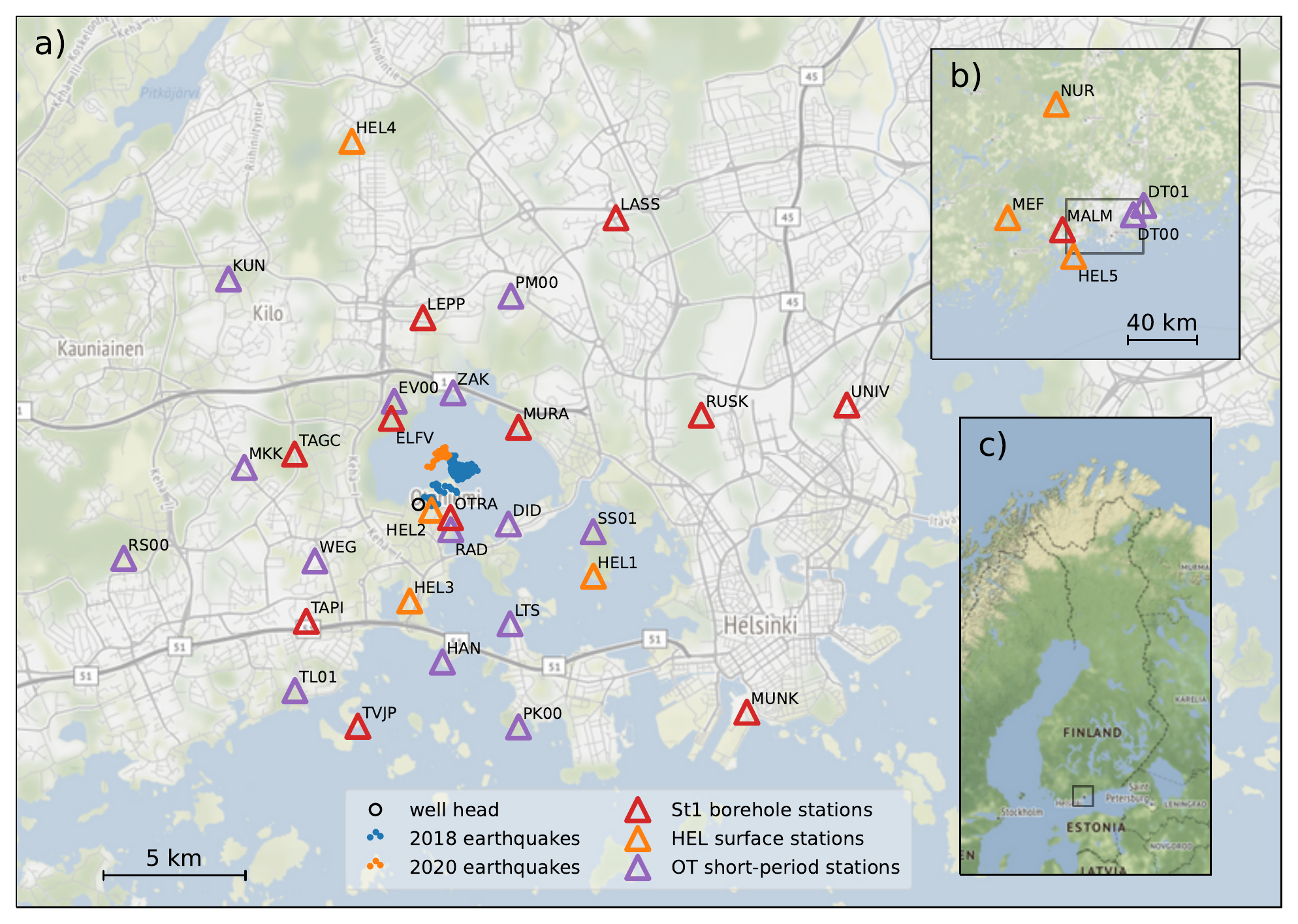}
\includegraphics[width=0.3\textwidth]{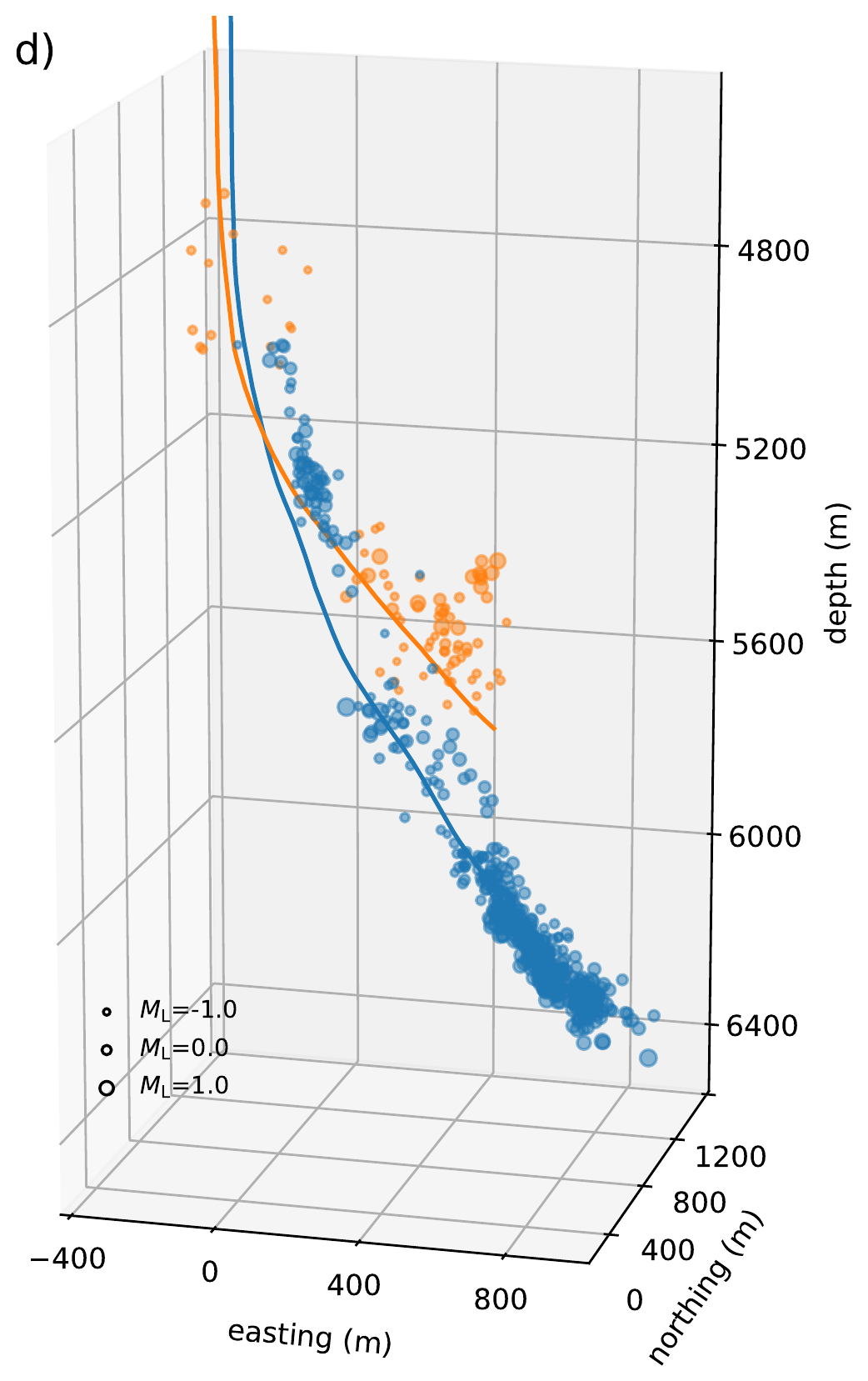}
\caption{
(a) Map of the Helsinki metropolitan area. Colored triangles indicate borehole and surface broadband and short-period stations. Blue and orange dots in the center show locations of the analyzed earthquakes induced during the 2018 and 2020 stimulations. The location of the well head is indicated.
(b) The larger scale map shows seismic stations outside the region displayed in map~(a). The rectangle indicates the area of map~(a).
(c) Map of Finland and neighboring countries. The rectangle indicates the area of map~(b).
(d) Trajectories of the 2018 and 2020 drill holes together with induced seismicity. Northing and easting are relative to the location of the well head. This illustration contains 484 events induced in 2018 and 86 events induced in 2020 \citep{HELdata2018IMS,HELdata2020ISUH}.
}
\label{fig:map}
\end{figure}

In this study we analyze earthquake data from the ${\sim}\SI{6}{km}$ deep stimulation experiments in the Helsinki, Finland, metropolitan area, that were performed in 2018 and 2020 using two sub-parallel wells to establish a geothermal district heating system (Figure~\ref{fig:map}) \citep{Kwiatek2019,Leonhardt2020}.
We use envelopes of seismograms from 233 events in 2018 and 22 events in 2020 in the {\Ml}0.0 to {\Ml}1.8 range that were recorded by local networks consisting of permanent and temporary borehole and surface stations \citep{Kwiatek2019,Hillers2020,Rintamaeki2021}.
The study area is located in the southern Fennoscandian Shield, where the level of natural background seismicity is relatively low \citep{Veikkolainen2021}.
The Finnish National Seismic Network detects between 10000 and 20000 seismic events per year. However, only a few hundred are earthquakes located in its own or neighboring territory with magnitudes that rarely exceed {\Ml}2.5; the majority of the detections are associated with explosions related to mining or infrastructure development.
In this seismically relatively quiet cratonic environment induced earthquake signals are thus also essential to update information of medium properties.
Erosion processes associated with glaciation and postglacial uplift stripped away sedimentary deposits, and the bedrock environment leads to good-quality signals of the small magnitude events.
\par
This facilitates our estimates of source, medium, and site parameters from seismogram envelopes that here have a duration of about \SI{20}{s}.
The energy levels in the direct $S$~wave and the scattered $S$~coda wavefield
are modeled using the theory of radiative transfer \citep{Sato2012}.
Matching observed and radiative transfer based synthetic envelope shapes constrains four different frequency dependent functions. 
These include average scattering and inelastic attenuation, a site term for each receiver location, and the displacement source spectrum for each event.
We refer to this method as Qopen which acronymizes ``separation of intrinsic and scattering \underline{$Q$} by
envel\underline{ope} inversio\underline{n}'',
and which is implemented in the Qopen software package \citep{Eulenfeld2016}.
This envelope method has been applied to other compact event sequences in different tectonic environments including the Molasse basin and the upper Rhine rift in Germany \citep{Eulenfeld2016}, west Bohemia in the German-Czech border region \citep{Eulenfeld2021}, but also to USArray data with a focus on the spatial variability in the attenuation properties \citep{Eulenfeld2017}.
\par
Alternative coda analysis methods to constrain source and medium effects equally consider narrow-band envelopes \citep[and others]{Mayeda2003, Holt2021}, but the advantages of the radiative transfer model are that it depends explicitly on the scattering medium properties, and resulting magnitudes do not require any calibration.
The shapes of the long seismograms constrain intrinsic or inelastic attenuation, in contrast to methods targeting the decay of the direct wave amplitude \citep{Boatwright1978, Anderson1984}.
Therefore, both attenuation mechanisms can be inverted for separately, similar to multiple lapse time window analysis \citep{Fehler1992, Hoshiba1993}.
Yet other established approaches to separate source, path, and site effects include the generalized inverse technique \citep{Andrews1986,Castro1990,Parolai2000},
iterative event-stacking schemes \citep{Prieto2004,Shearer2006,Yang2009,Trugman2020},
and the empirical Green's function method \citep{Berckhemer1962,Mueller1985,Abercrombie2015}.
These methods underpin the study of earthquake source parameters that are then obtained by the application of spectral source models to the isolated source spectra \citep{Brune1970,Boatwright1978,Abercrombie1995}.
Scaling relations between these and further derived parameters such as stress drop are essential for our understanding of the physical source processes, and they find application in ground motion prediction and hazard assessment scenarios.
The discrepancy between source parameters and scaling relations estimated with different approaches from the same data \citep{Kaneko2014,Shearer2019,Abercrombie2021} 
highlights the need for investigations into robust source spectra determination,
and into the sensitivity of model parameters to different processing chains and choices.
\par
The key features of our study---compact source region, good network, hard rock environment, high-quality signals, radiative transfer Green's function-based envelope analysis, wide \SI{3}{Hz} to \SI{200}{Hz} frequency range---yield diverse results on attenuation properties, site effects, and earthquake scaling relations that demonstrate the significance of the Qopen method
for isolating inconsistencies based on different models. Main observations associated with the four target functions that have been obtained using the larger 2018 data set include first frequency dependent scattering $Q^{-1}\sca$ values between $10^{-3}$ and $10^{-4}$ that are compatible with the results obtained by 20 other studies in different environments.
Second, compared to these 20 studies the
intrinsic attenuation $Q^{-1}\intr$ is with ${\sim}\num{e-2}$ relatively high below \SI{5}{Hz}, but plunges to extremely low values smaller than $10^{-4}$ at frequencies above  \SI{7}{Hz}.
Third, the site effect terms show interesting amplification at high frequencies above ${\sim}\SI{50}{Hz}$ with implications for ground motion studies in the Fennoscandian Shield.
Fourth, the employed spectral source model \citep{Boatwright1978} yields high frequency falloff rates between 1.4 and 2.2,
and corner frequencies that scale with moment as $M_0{\propto}f_c^{-5.37\pm0.24}$, which differs from the $M_0{\propto}f_c^{-3}$ scaling associated with self-similarity. 
Using the values of the attenuation and site terms obtained with the 2018 data, we analyze the 2020 data in a retrospective monitoring mode to demonstrate the application of the Qopen software as real-time monitoring tool.
The 2020 data set is smaller, yet it resolves systematically lower stress drop estimates compared to the 2018 data, which we discuss in terms of variable slip speeds and medium changes in the reservoir.
\par
In the next Section~\ref{sec:data} we introduce the geothermal stimulation experiment, the seismic network, and the data.
The Method Section~\ref{sec:method} consists of two parts. We first provide an extended introduction of the theoretical basis and the assumptions of the employed radiative transfer-based envelope method
before we reproduce essentials of the technical implementation.
The Results Section~\ref{sec:results} and the Discussion Section~\ref{sec:disc} both separate the discussion of scattering and intrinsic attenuation, site effects, and source spectra, and the derived source parameters and scaling relations.

\section{Seismic network and data}
\label{sec:data}
We study data from the enhanced geothermal system (EGS) at about \SI{6}{km} below the Aalto University campus in the Otaniemi district of Espoo in the Helsinki metropolitan area (Figure~\ref{fig:map}) that has been developed by the St1 Deep Heat Oy company \citep{Kwiatek2019}.
During two weeks-long stimulations in 2018 and 2020 a reservoir fracture network was created between two wells by pumping high pressure freshwater into the crystalline Precambrian rock formation to enhance the fluid flow for an efficient heat exchange.
The first well is near-vertical down to \SI{4.7}{km} depth, and the north-east striking open hole section dips at about 40$^\circ$ down to \SI{6.1}{km} depth.
The open hole section of the second well is offset by about \SI{400}{m} to the north-west \citep{Leonhardt2020,Kwiatek2022}.
The stimulation through the first well lasted from 4~June 2018 to 22~July 2018,
and through the second well from 6~May 2020 to 24~May 2020.
Both stimulations employed multiple stages with adapted protocols to allow the induced energy to dissipate.
The first stimulation lasted longer (49~days compared to 19~days), used higher peak well-head pressures (\SI{90}{MPa} compared to \SI{70}{MPa}), and injected a larger volume of water (\SI{18000}{m^3} compared to \SI{2900}{m^3}).
In contrast to the sequence of events in Basel, Switzerland \citep{Haering2008}, Pohang, South Korea \citep{Ellsworth2019}, and Strasbourg, France \citep{Schmittbuhl2021},
that all led to the termination of the associated EGS developments, the ground shaking around Otaniemi did not exceed the limit set by local authorities \citep{Ader2019}.
The largest induced event was {\Ml}1.8, the traffic light system red-level operation stop size was set to {\Mw}2.0.
\par
The stimulations activated an existing fracture network and induced many thousands of events in compact zones around and between the two wells (Figure~\ref{fig:map}d) \citep{Kwiatek2019,Leonhardt2020,Kwiatek2022}.
Three distinct seismicity clusters developed during the 2018 experiment that extended approximately \SI{1400}{m} in depth and \SI{1100}{m}$\times$\SI{500}{m} laterally.
The distance between the 2020 and 2018 seismicity is smaller than the dimension of the 2018 clusters \citep{Kwiatek2022}. 
Because of this compact source region we can expect that our results are not controlled by systematic structural complexities between different parts of the reservoir.
We tend to attribute observed variations in earthquake properties to changes associated with the fluid injections, although heterogeneities associated with damaged and permeable zones \citep{Kwiatek2019} can also modulate earthquake behavior.
\par
The St1 company established 12 borehole satellite stations sampling at \SI{500}{Hz} between \SI{240}{m} (TAPI) and \SI{1200}{m} (MURA) depth in the area (red symbols in Figures~\ref{fig:map}a, \ref{fig:map}b) for industrial and regulatory purposes. The continuous data are transmitted to the Institute of Seismology at the University of Helsinki (ISUH) as part of a monitoring agreement and can be used by ISUH for monitoring and research.
The horizontal orientation of the downhole instruments is not calibrated.
To complement the two nearest stations from the Finnish National Seismic Network MEF and NUR (Figure~\ref{fig:map}b),
ISUH has been deploying and operating eight permanent broadband stations in the Helsinki area. 
These stations sample at \SI{250}{Hz}, except for the MEF station that samples at \SI{100}{Hz} \citep{Rintamaeki2021}.
In addition, ISUH managed temporary seismic networks in 2018 and 2020. Each network consisted of about 100 instruments that were organized as stand-alone stations and in mini arrays \citep{Hillers2020,Rintamaeki2021}.
Here we consider short-period stations from the OT subnetwork that sampled at \SI{400}{Hz} \citep{Rintamaeki2021}.
All instruments record ground velocity.
The associated 3, 4, 7, or 25 station arrays were deployed, roughly, on a \SI{100}{m}$\times$\SI{100}{m} patch, but we use data from only one station per site.
From each array, we simply use the first sensor in the site specific name list, without further quality control.
We process data from 36 stations of which one, PVF, ends up not being used in the analysis due to its $\sim$\SI{70}{km} distance to the source region \citep{Hillers2020}.
\par
The 2018 network has a good overall azimuthal coverage which supports the network averages of source and medium effects.
Most stations are located within a \SI{10}{km} radius around the stimulation site.
This short-distance range does not complicate our attenuation analysis, since the $Q$ estimates are constrained by the time dependent coda decay, and not by the distance dependent amplitude of the direct $S$~wave.
The 2020 network occupied more sites \citep{Rintamaeki2021}, but in our monitoring-mode application we use stations from only those locations that were also occupied during the 2018 stimulation.
\par
\begin{SCfigure} \centering
\includegraphics[width=0.5\textwidth]{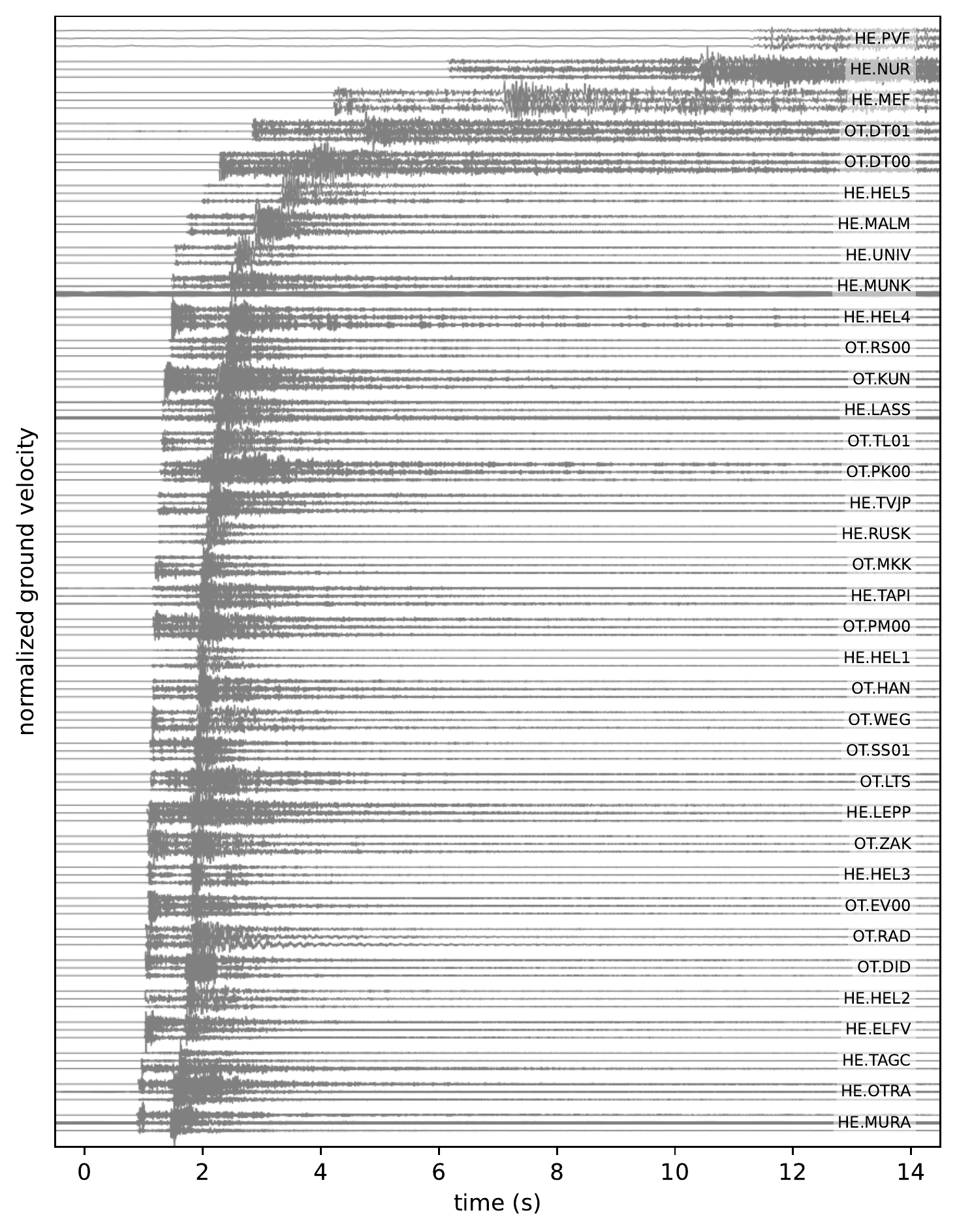}
\caption{Three-component waveforms for a {\Ml}1.6 earthquake in 2018. Time is relative to the origin time. For each station one trace is displayed for each component (ZNE from top to bottom). Data are \SI{1}{Hz} high-pass filtered and normalized.
\newline\newline
}
\label{fig:rawdata}
\end{SCfigure}

The absence of a sedimentary layer in the geological setting of the old and cold Fennoscandian Shield means the surface instruments sat on the same outcropping bedrock units that were stimulated at depth.
Precambrian rocks overlying the ${\sim}\SI{50}{km}$ deep Moho discontinuity are characterized by high seismic velocities with $v\ind P$ and $v\ind S$ values close to the surface around \SI{6}{km/s} and \SI{3.5}{km/s} \citep{Tiira2020}.
A regional average passive surface wave dispersion analysis resolved a few tens of meters thick low-velocity layer in the area \citep{Hillers2020} associated with weathered rock and topographic depressions filled with unconsolidated post-glacial sediments.
Figure~\ref{fig:rawdata} displays high-pass filtered seismograms of a {\Ml}1.6 event.
The good $S$~wave signal-to-noise ratio facilitates the derivation of rotational ground motion from the translational mini-array data \citep{Taylor2021}.
We use records of 233 events induced in 2018 
\citep{HELdata2018IMS} that include the same 204 events analyzed by \citet{Taylor2021},
together with 22 events induced during the 2020 stimulation \citep{HELdata2020ISUH},
all with sizes in the Finnish local magnitude range {\Ml}0.0 to {\Ml}1.8.
The routine detection and analysis procedure is summarized in \citet{Hillers2020} and \citet{Rintamaeki2021}. 
Magnitudes {\Ml} are estimated from manually revised amplitudes \citep{Uski1996,HELdata2018ISUH}. 
We use updated picks and event locations for the 2018 events \citep{Gal2021,HELdata2018IMS}.
Differences between the locations shown in Figure~\ref{fig:map} and the results by \citet{Kwiatek2022} and \citet{Leonhardt2020} are associated with different sensor configurations, velocity models, and processing techniques, but are not relevant for our conclusions here.

\section{Method}
\label{sec:method}

In this section we summarize the theoretical concepts and the implementation of the employed Green's function envelope modeling using radiative transfer \citep{SensSchoenfelder2006a,Eulenfeld2016,Eulenfeld2017,Eulenfeld2021}.
Readers familiar with these developments may prefer to proceed to Section \ref{sec:results}.

\subsection{Relation to other methods, theoretical concepts, and assumptions}
\label{sec:methodbasics}
Qopen estimates the regional average scattering strength $g$ and intrinsic attenuation $b$, the spectral source energy $W$ for each event, and a site term $R$. All four terms are functions of frequency.
The method complements an array of techniques that isolate and constrain contributions from source, path, and site by averaging records of many earthquakes observed at multiple stations that ideally cover a diverse azimuth and distance range.
The Qopen narrow-band processing and the assemblage of source and site spectra is similar to the generalized inverse technique \citep{Castro1990}, which, too, fixes the average site term to unity \citep{Picozzi2022}, and which also allows estimates of the seismic moment by the application of an earthquake source model.
The coda calibration tool \citep{Mayeda2003} also employs narrow-band coda envelopes to calculate source spectra, but in contrast to the physics-based Qopen model the empirical parameterization of the coda envelope shape requires the calibration to independently estimated source spectra.
Regional iterative event-stacking schemes are a third alternative to separate source, propagation, and site contributions from spectra of short-duration $P$ and $S$~wave signals \citep{Prieto2004,Shearer2006,Yang2009,Shearer2019,Trugman2020,Abercrombie2021}.
These involve an arbitrary scaling associated with an empirical reference spectrum, which prohibits magnitude and associated source parameter estimates such as stress drop or energy release.
In these studies, earthquake size is constrained using independent observations,
which is generally achieved by tying the seismic moment magnitudes to local magnitude estimates.
A fourth approach is the empirical Green's function method,
which removes path and site effects from the observed spectra of a larger target event by deconvolving the signal of a suitable smaller event.
These empirical Green's functions can be spectra from single events \citep{Berckhemer1962,Mueller1985,Abercrombie2015} or stacks \citep{Ross2016,Ruhl2017}.
Note that these latter two event-stacking and empirical Green's function methods target $P$ or $S$~wave pulses that are much shorter compared to the envelope duration considered here.

\par
The here employed envelope method simultaneously constrains the parameters $g$ and $b$ that are proportional to scattering $Q\sca^{-1}$ and intrinsic $Q\intr^{-1}$ of shear waves

\begin{align}
  Q^{-1}\sca &= \frac{g v\ind S}{2\pi f} & Q^{-1}\intr &= \frac b{2\pi f} & Q^{-1} = Q^{-1}\sca + Q^{-1}\intr,
  \label{eq:Q}
\end{align}

where $v\ind S$ is the average $S$~wave velocity in the sampled volume, and $Q^{-1}$ is the total attenuation.
Frequency is denoted by $f$, and the analysis is applied to narrow-band filtered seismogram envelopes, where the 13 frequencies are equally spaced between 3~Hz and 200~Hz on the logarithmic scale.
The spectral source energy estimate $W\fof f$ for a single event is converted to the $S$~wave source displacement spectrum $\omM\fof f$, where $M$ denotes the Fourier transform of the seismic moment time function.
The earthquake source spectra are compiled from the individual observations over the full frequency range, which results in relatively smooth shapes that are then used to constrain spectral source models, such as the Brune model \citep{Brune1970} or its alternatives.
\par
At first, $g$ and $b$ are estimated together with $W$ and $R$, after which estimates of $W$ and $R$ are updated using the fixed values for the medium parameters $g$ and $b$. This step eliminates possible trade-offs between $b$, $g$ and $W$, $R$.
The perfect trade-off between source term $W$ and site term $R$ is approached by fixing average site effect values arbitrarily at all stations or at a subset of reference stations to unity, similar to some of the reviewed methods.
\par
The Qopen approach to constrain $Q\sca^{-1}(f)$, $Q\intr^{-1}(f)$, $W\fof f$, and $R\fof f$ differs from these methods,
since it is based on the Green's function modeling of the direct $S$~wave energy and the coda decay properties using radiative transfer.
The synthetic seismogram envelopes account for isotropic scattering, intrinsic attenuation, and geometrical spreading. 
The method can separate and resolve the two scattering and intrinsic attenuation components because of their different effects along the envelope, as energy is scattered away from the direct $S$~wave and redistributed to later arriving coda waves. 
Scattering attenuation governs the shape of the envelope, i.e., the relative level of energy density in different parts of the envelope. 
In contrast, intrinsic attenuation controls the exponential decay of energy density with time.
\par
The Green's function model is based on radiative transfer, a framework that describes the propagation of energy density through heterogeneous, scattering media, that was first developed for light scattering \citep{Chandrasekhar1950} and then adopted to seismic wave scattering \citep{Wu1985,Margerin2005,Sato2012}.
Common to other applications of randomized seismic wavefields the theory assumes the equivalence of ensemble average and time average, i.e., a sequence of scattered seismic wave propagation is a single realization of an ergodic process.
The medium properties described by the parameters $g$ and $b$ can vary spatially, but in our application the compact scale of the target area around the reservoir suggests to assume constant half-space values of $g$ and $b$ in this study, compared to the large-scale USArray application of \citet{Eulenfeld2017}. We use additional information on average medium properties including shear wave velocity and density, which are typically available with enough accuracy to support the results. The underlying model describes isotropic scattering---an assumption which is presumably not accurate for the broad frequency range analyzed in this study.
The discrepancy can be resolved by interpreting the scattering strength as transport variable, which means it is not so much a proxy for the average number of individual scattering events, but more an indicator of the medium efficiency to redistribute seismic energy to different directions associated with multiple scattering \citep[their Figure~8]{Gaebler2015, Margerin2016}. 
The only part of the shear wave envelope which is not adequately modeled with this approach is energy around the arrival of the direct $S$~wave, because multiply forward scattered waves lead to an envelope shape which cannot be described by the used Green's function for isotropic scattering. This is why the envelope in a short window around the $S$~wave arrival is time averaged \citep{SensSchoenfelder2006a, Eulenfeld2016}, similar to the approach of multiple lapse time window analysis used to separate intrinsic attenuation and scattering attenuation \citep{Fehler1992, Hoshiba1993}. For a comparison of Qopen and the multiple lapse time window analysis we refer to \citet{Laaten2021}.

\subsection{Implementation of the Qopen method}
\label{sec:methodqopen}

In this section we detail the implementation of the three key iterative steps of the Qopen algorithm, which include, one, estimating intrinsic $g$ and scattering $b$ attenuation using a subset of the strongest signals,
two, solving for source spectra and site terms $W$ and $R$ with fixed $g$ and $b$,
and three, solve again for $W$ with $g$, $b$, $R$ fixed to the average values determined in the previous steps \citep{Eulenfeld2021}.
\par
The observed energy density envelopes $E\obs$ are calculated from the three-component instrument corrected and filtered seismic velocity records $\dot u_c$ using the Hilbert transform $\mathcal H$ (\citealp{Sato2012}, page~41; \citealp{Eulenfeld2016}, equations 3--4)

\begin{equation}
E\obs\fof{t, \vec r} = \frac{\rho\sum_{c=1}^3\parent{\dot u_c\fof{t, \vec r}^2 + \mathcal H\fof{\dot u_c\fof{t, \vec r}}^2}}{2C\ind{energy}\Delta f},
\label{eq:Eobs}
\end{equation}

with the mean mass density $\rho$, energy free surface correction $C\ind{energy}{=}4$ \citep{Emoto2010}, filter bandwidth $\Delta f$ \citep[e.g.][]{Wegler2006a}, and lapse time $t$. The vector $\vec r$ is pointing from the earthquake location to the station location. Station MUNK shows a corrupt E~component for a small fraction of the earthquakes (Figure~\ref{fig:rawdata}). 
We decide to use the N~component data instead in Equation~\ref{eq:Eobs}, i.e., exceptionally we use ZNN instead of ZNE data.
The central frequencies of the Butterworth bandpass filter are taken from the range between \SI{3}{Hz} and \SI{192}{Hz}; the filter width is $2/3$ of the central frequency. The filter corresponding to the maximum central frequency \SI{192}{Hz} is a Butterworth highpass filter with cut-off frequency \SI{128}{Hz}. The filters have two corners and are applied forward and backward for zero phase shift.
The narrow-band envelopes are smoothed with a \SI{1}{s} long central moving average.
\par

\begin{figure}
\centering
\includegraphics[width=0.49\textwidth]{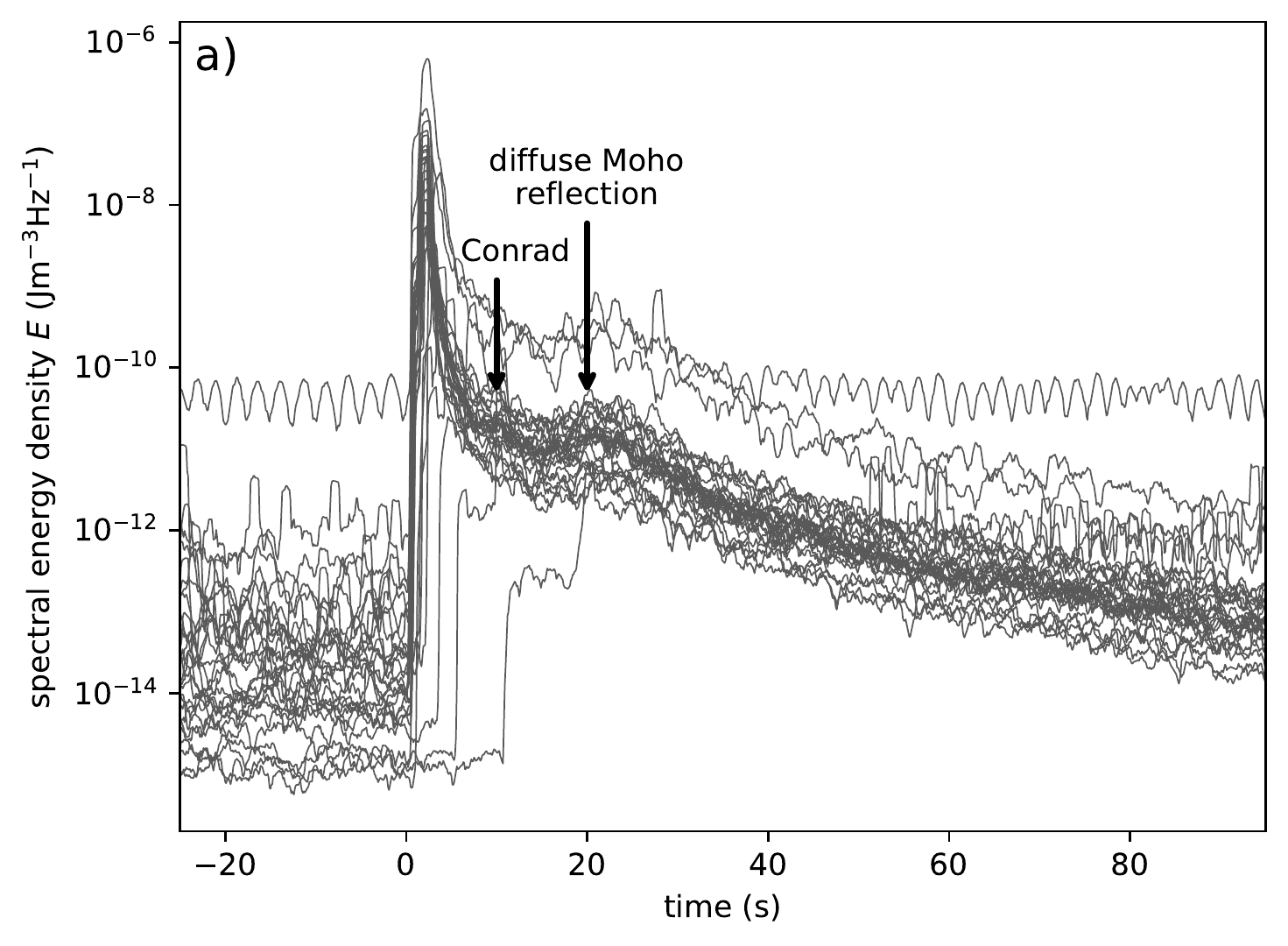}
\includegraphics[width=0.49\textwidth]{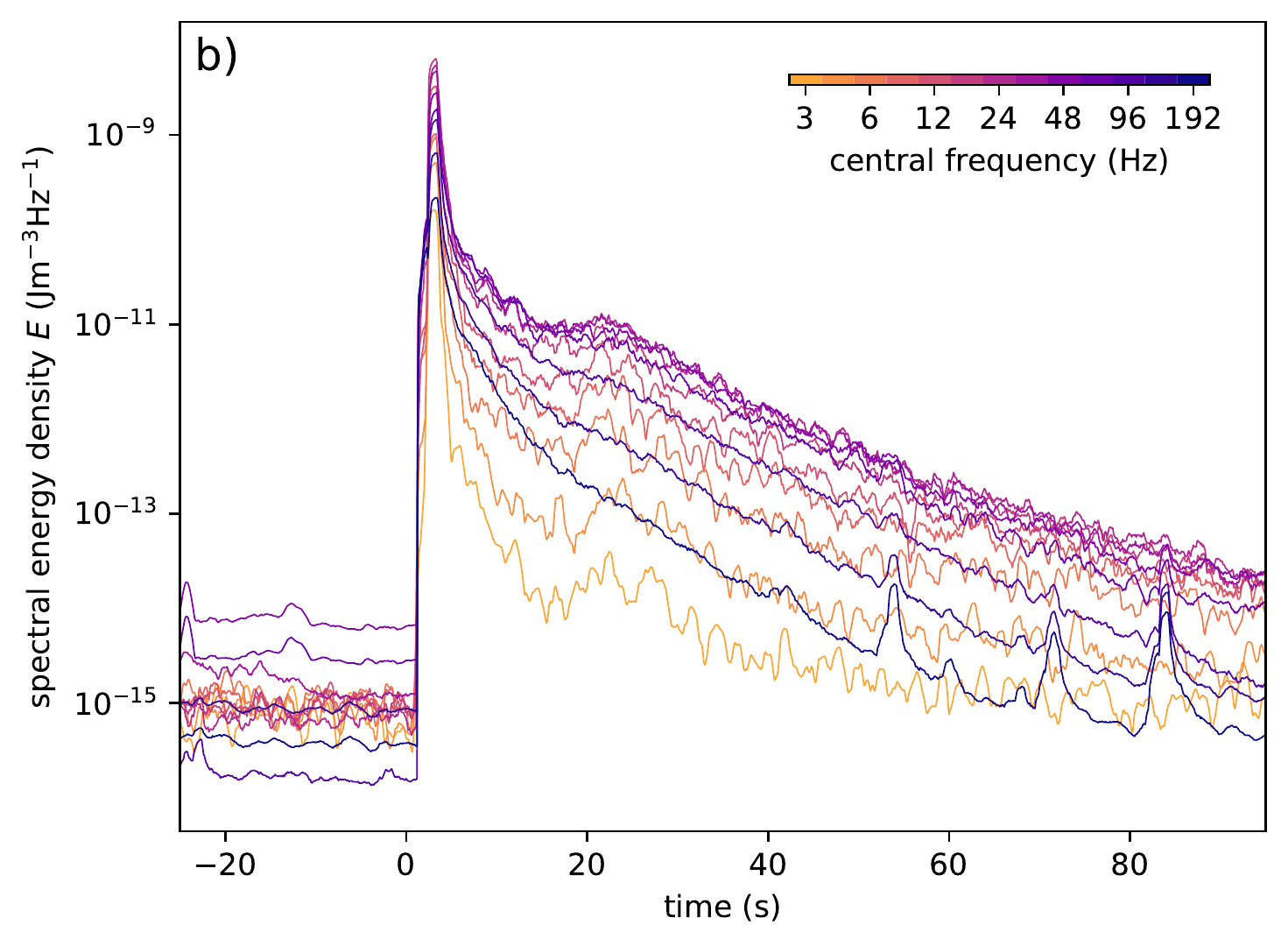}
\caption{(a) Observed smoothed envelopes of spectral energy density for each of the used 36 stations in the \SI{8}{Hz} to \SI{16}{Hz} frequency range for a {\Ml}1.6 earthquake.
A diffuse Moho reflection is visible at around \SI{20}{s}. A weaker transient increase in the energy envelopes at \SI{10}{s} likely corresponds to a diffuse reflection at the mid-crustal Conrad discontinuity.
(b) Smoothed envelopes for data from the same event as in (a) recorded at a single borehole station (MALM) in different frequency bands indicated by different colors. The Moho reflection is visible for frequencies up to \SI{100}{Hz}.
}
\label{fig:data}
\end{figure}

Figure~\ref{fig:data}a displays energy envelopes in the frequency band \SI{8}{Hz} to \SI{16}{Hz} observed at different stations for a single earthquake.
They feature the typical sharp increase associated with the arrival of the ballistic wave, and the indicative decay of the scattered coda wave energy that approaches the pre-event noise level after tens of seconds propagation time. 
An increase in the coda envelopes for all stations is visible between \SI{18}{s} and \SI{28}{s}. The observed 1.5-fold to 2-fold increase for the displayed frequency band is interpreted as diffuse Moho reflection (Section~\ref{sec:disc}) and is visible up to \SI{100}{Hz} (Figure~\ref{fig:data}b).
For each frequency, the observed energy densities are compared to synthetic envelopes which are given by \citep{Eulenfeld2016}

\begin{equation}
E\md\fof{t, \vec r} = WR\fof{\vec r}G\fof{t, \vec r, g}e^{-bt} \,, \label{eq:Emod}
\end{equation}

where $W$ is the spectral source energy of the earthquake, and $R\fof{\vec r}$ is the energy site term at a station. The Green's function $G\fof{t, \vec r, g}$ with scattering strength $g$ accounts for the direct wave and the scattered wavefield and is given by the approximation of the solution for three-dimensional isotropic radiative transfer \citep{Paasschens1997}.
This Green's function modeling yields moment estimates that agree well with
independently obtained results \citep{Eulenfeld2021}.
The exponential term $e^{-bt}$ describes the intrinsic damping with time and depends on the absorption parameter $b$.
Performing the inversion separately for different frequencies yields the $f$ dependence of $W$, $R$, $g$, and $b$.
\par
The spectral source energy $W\fof f$ is converted to the $S$~wave source displacement spectrum $\omMf$ using Equation~11 of \citet{Eulenfeld2016} \citep[page~188]{Sato2012}

\begin{equation}
  \omM\fof f = \sqrt{\frac{5 \rho v\ind S^5 W\fof f}{2\pi f^2}} \,.\label{eq:sds}
\end{equation}

Note that the spectral source energy $W\fof f$ controls the energy envelope level (Equation~\ref{eq:Emod}).
The displacement spectrum and the envelope properties are thus linked by a framework that synthesizes seismogram envelopes of realistic earthquake sources in an inhomogeneous medium \citep{Sato2012},
in contrast to EGF studies that link source spectra to amplitudes of ballistic waves.
In Equation~\ref{eq:sds} the density is $\rho{=}\SI{2700}{kg/m^3}$ and the wave speed is $v\ind S{=}\SI{3.5}{km/s}$ \citep{Tiira2020}.
The source displacement spectrum can be fitted by a general source model of the form 

\begin{equation}
  \omMf = M_0 \parent{1+\parent{\frac f{f\ind c}}^{\gamma n}}^{-\frac 1\gamma}, 
\label{eq:sourcemodel}
\end{equation}

with seismic moment $M_0$ and corner frequency $f\ind c$ \citep{Boatwright1978}. 
Note that the numerator in this expression does not involve an attenuation term, as this is accounted for by the $g$ and $b$ parameters.
The shape parameter $\gamma$ describes the sharpness of the transition between the flat low-frequency level and the high-frequency amplitude falloff that scales $f^{-n}$,
and fixing $n{=}2$ yields the so-called omega-square spectral source displacement model.
The shape parameter values $\gamma{=}1$ and $\gamma{=}2$ correspond to the widely used \citet{Brune1970} and \citet{Boatwright1980} models.
Earthquake source spectra studies have explored the systematic differences in scaling relations obtained with the Brune-type or the Boatwright-type model,
but the two approaches yield overall similar results if the corner frequency is well within the center of the analyzed frequency range \citep{Kaneko2014,Abercrombie2016,Shearer2019,Abercrombie2021}.
Here we use $\gamma{=}2$. 
\par
To estimate source parameters from the observed source spectra we use a nonlinear least squares fitting method from the SciPy ecosystem that employs the L-BFGS-B algorithm \citep{Zhu1997}.
Qopen employs this method on the logarithm of Equation~\ref{eq:sourcemodel} and automatically down-weights outliers \citep{Eulenfeld2016}.
The equidistant logarithmic frequency sampling leads to model parameter estimates that are not biased by an oversampling of high frequencies.
A frequency dependent weighting as employed in some EGF studies is not needed \citep{Kaneko2014}.
\par
For each frequency, the number of equations in the overdetermined system
\begin{equation}
\ln E\obs\fof{t, \vec r} = \ln E\md\fof{t, \vec r}  \label{eq:system}
\end{equation}
is large. It is the number of samples in the coda plus one average direct wave datum summed over all stations. The number of variables is the sum of the number of stations plus two if each event is inverted separately as in this study \citep{Eulenfeld2016}.
We use a direct $S$~wave window (\SI{-0.5}{s}, \SI{3}{s}) relative to the $S$ onset. The coda window starts at the end of the direct wave window and ends \SI{18}{s} after the origin time to exclude the transient increase in spectral energy associated with the Moho reflection.
The coda window is shorter if the signal-to-noise ratio (SNR) drops below 2 or if energy increases again due to occasional transients, but the window has to be at least \SI{5}{s} long for data to be included in the analysis.
The noise level is estimated in a \SI{30}{s} long time window preceding the origin time and it is removed from the envelope.
The envelope in the direct $S$~wave window is averaged to mitigate the effect of forward scattering \citep{Eulenfeld2016}. This average is weighted according to the length of the direct $S$~wave window.
\par

\begin{figure}
\centering
\includegraphics[width=0.9\textwidth]{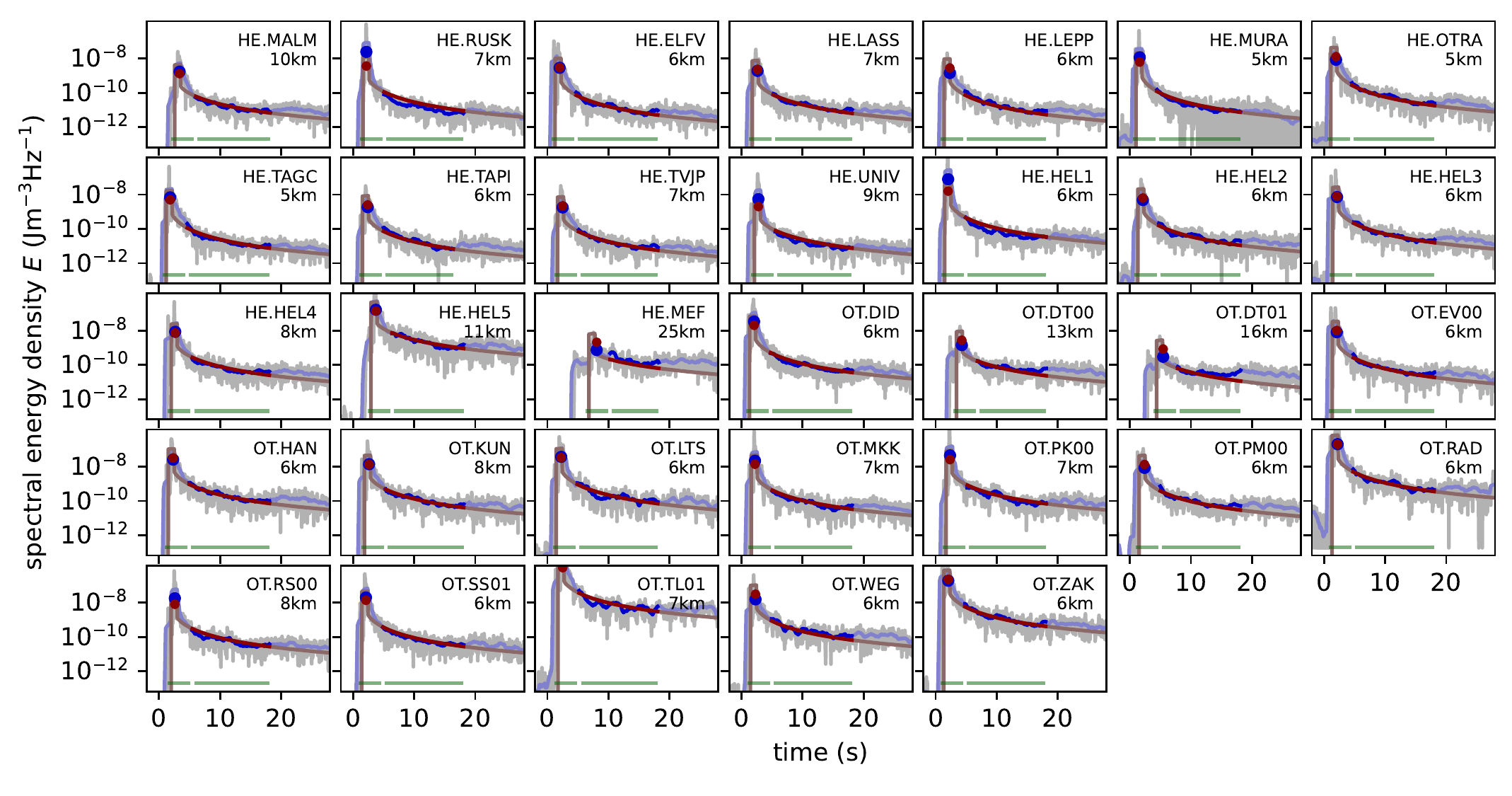}
\caption{Observed and synthetic envelopes of spectral energy density for each station in a separate panel in the \SI{16}{Hz} to \SI{32}{Hz} frequency range calculated with Qopen for a {\Ml}1.6 earthquake in 2018.
Observed envelopes are displayed with grey lines, smoothed observed envelopes with blue lines, and synthetic envelopes with red lines. The green bars at the bottom indicate the two time windows used in the inversion. Envelopes in the direct $S$~wave time window indicated by the blue and red dots are averaged, envelopes in the coda time window are displayed as dark blue and dark red lines. Coda time windows that are shorter than \SI{5}{s} are not included in the analysis.
}
\label{fig:fits}
 \vspace{0.5cm}
\includegraphics[width=0.9\textwidth]{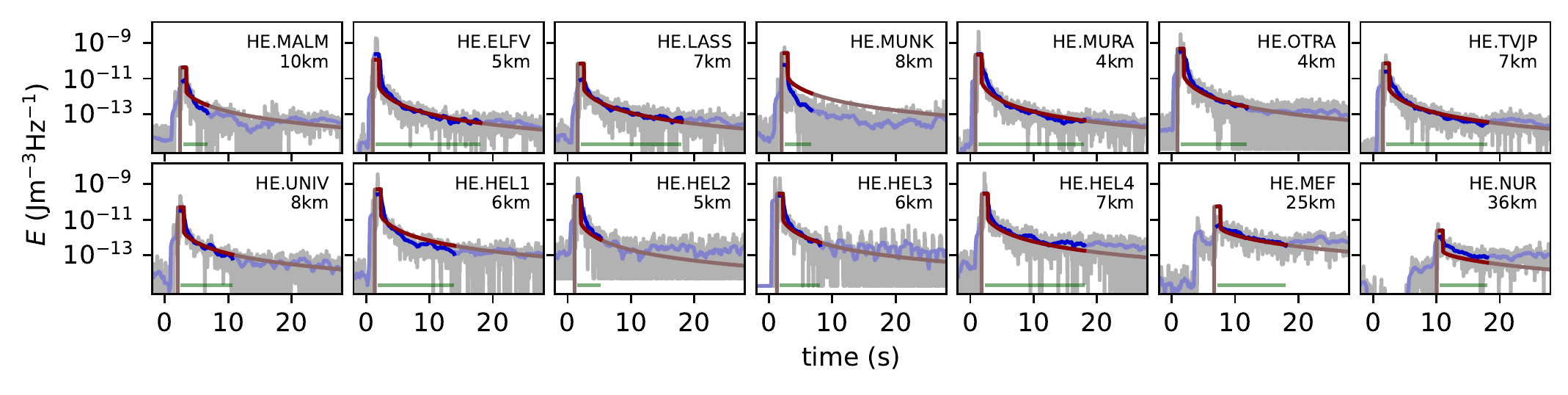}
\caption{Envelope fits for a {\Ml}0.8 2020 earthquake in the \SI{16}{Hz} to \SI{32}{Hz} frequency range obtained with Qopen operating in monitoring mode. Because the focus is on the determination of a single constant $W$, the spectral source energy, it is feasible to use a single time window indicated by the green bar which starts at the theoretical $S$~wave travel time. Time windows that are shorter than \SI{2}{s} are not included in the analysis.}
\label{fig:fits2}
\end{figure}

In summary, the inversion consists of four main steps \citep[their Figure~3]{Eulenfeld2021}:

\begin{enumerate}
\item Using the 2018 data, intrinsic and scattering attenuation is estimated by solving equation system (\ref{eq:system}) for $g$, $b$, $W$, and $R$ for all frequency bands and all earthquakes separately. An example is illustrated in Figure~\ref{fig:fits}. In each frequency band $g$ and $b$ are geometrically averaged over different events \citep{Eulenfeld2016} and converted to $Q$ values using Equation~\ref{eq:Q}. For this step we use all 36 earthquakes with $M_L{\geq}1$.
\label{step1}
\item Site terms are obtained by again solving the system of equations (\ref{eq:system}) for $W$, $R$ with fixed medium parameters $g$ and $b$ for all frequency bands and for the earthquakes used in step~\ref{step1} separately. 
Because of the co-linearity of $R$ and $W$ in Equation~\ref{eq:Emod} and because each earthquake might be registered at a different set of stations, the site terms $R$ are re-aligned as described in Section~2.2 of \citet{Eulenfeld2017} to ensure self-consistent observations of site terms. For reference the geometric mean of borehole stations MALM and RUSK is fixed to 0.25 for all frequencies, because stations MALM and RUSK with sensor depths around \SI{300}{m} show the lowest amplification over the full frequency range. The chosen value of 0.25 for borehole stations corresponds to a neutral unit amplification including a compensation for the earlier applied free surface correction in Equation~\ref{eq:Eobs}. For this and the following steps the coda time window has to have a minimum duration of \SI{2}{s} to be included in the analysis.
\item The source displacement spectra $\omM\fof f$ are compiled by solving the system of equations (\ref{eq:system}) a third time for $W$ with fixed parameters $R$, $g$, and $b$ for all frequencies and all earthquakes with $\Ml{\geq}0$ separately.
Finally, spectral source energies are converted to source displacement spectra using Equation~\ref{eq:sds} and source parameters are determined by fitting the source model Equation~\ref{eq:sourcemodel} for all earthquakes with results in more than 4 frequency bands.
\item Additionally, we process the 2020 data using the setup and the parameters $R$, $g$, and $b$ estimated from the 2018 data to illustrate the utility of Qopen as monitoring software tool (Figure~\ref{fig:fits2}). Because attenuation parameters are already determined in step~\ref{step1}, the discrimination between direct $S$~wave window and coda window is not necessary. Therefore, we use a single time window that starts at the theoretical $S$ onset estimated from the distance and mean $S$~wave velocity, and the window ends \SI{18}{s} after the origin time or if the envelope reaches a SNR level of 2. This means no manual pick information are used in the processing of the 2020 data.
\label{step4}
\end{enumerate}

\section{Results}
\label{sec:results}

In this section we give a concise overview of relevant inversion results including observations of attenuation properties,
site terms, earthquake source spectra, earthquake source parameters, and scaling relations. 

\subsection{Scattering properties and intrinsic attenuation in the Fennoscandian Shield}
\label{sec:results_Q}

\begin{figure}
\centering
\includegraphics[width=0.5\textwidth]{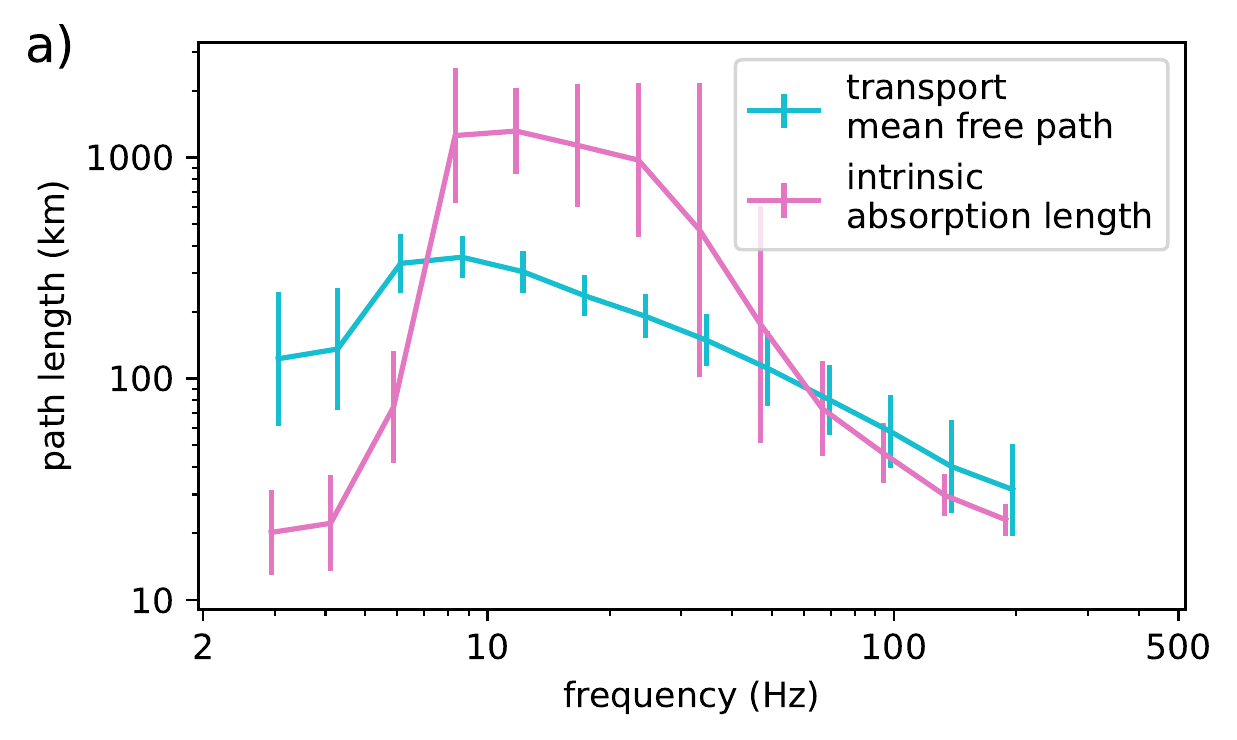}
\includegraphics[width=0.9\textwidth]{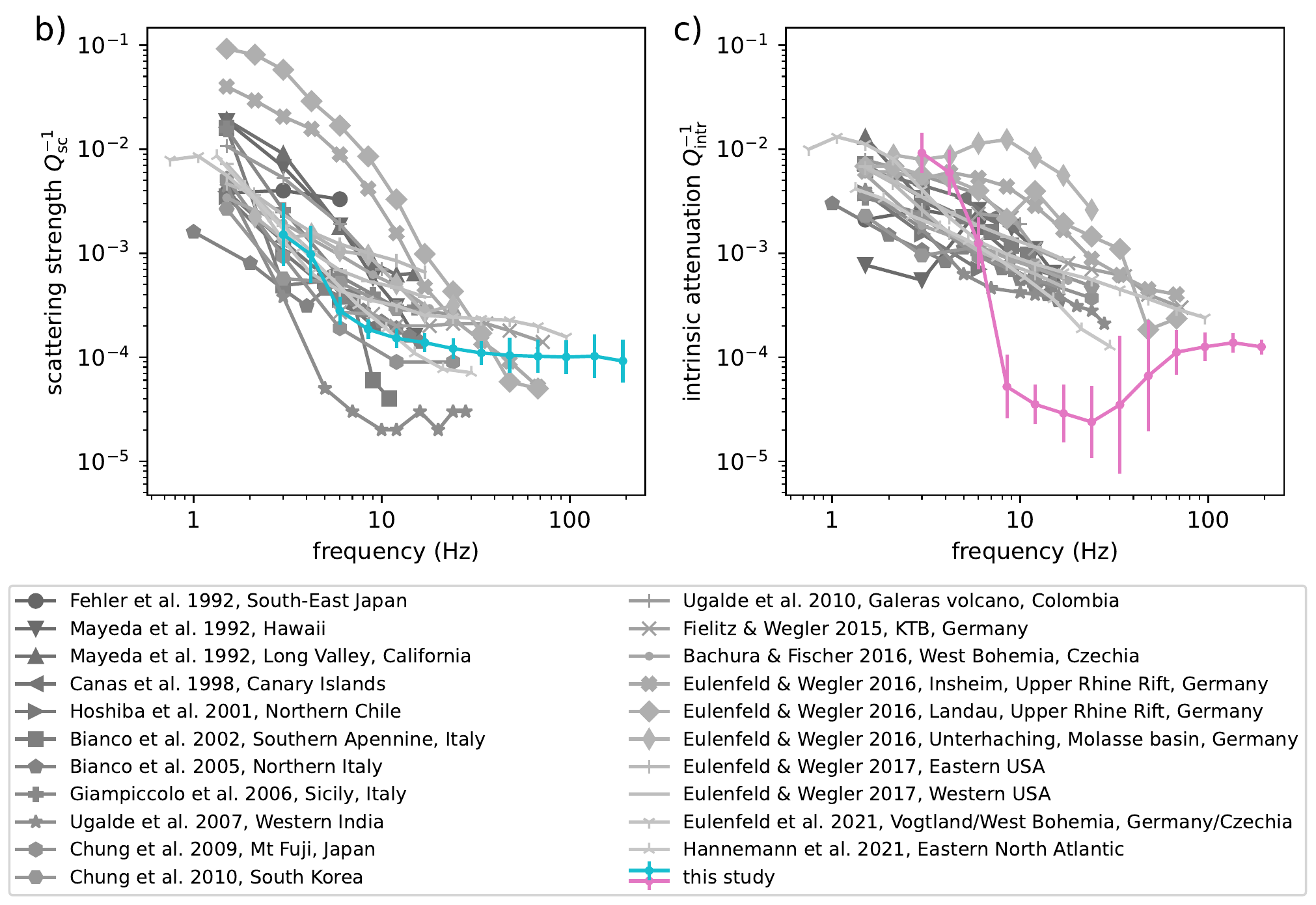}
\caption{(a)~Transport mean free path and absorption length of $S$~waves as a function of frequency. 
(b)~Scattering strength and (c)~intrinsic attenuation as a function of frequency compared to results from other studies.
}
\label{fig:Q}
\nocite{Fehler1992, Mayeda1992, Mayeda1992, Canas1998, Hoshiba2001, Bianco2002, Bianco2005, Giampiccolo2006, Ugalde2007, Chung2009, Chung2010, Ugalde2010, Fielitz2015, Bachura2016, Eulenfeld2016, Eulenfeld2016, Eulenfeld2016, Eulenfeld2017, Eulenfeld2017, Eulenfeld2021, Hannemann2021}
\end{figure}

Scattering attenuation is controlled by structural variability and heterogeneity in rock composition, fracturing, and porosity.
Intrinsic attenuation or dissipation is governed by viscous relaxation or internal frictional processes associated with boundaries along grains, cracks, fractures, and fluid movements.
Figure~\ref{fig:Q}a shows the transport mean free path and absorption length estimates in the Helsinki area that are obtained from the corresponding scattering and intrinsic absorption $Q^{-1}$ values shown in Figures~\ref{fig:Q}b and \ref{fig:Q}c
together with observations from various other tectonic environments for comparison. 
We can collate the results in the same figure because they all are obtained with Qopen or a similar method, the multiple lapse time window analysis.
Whereas the wavefields excited by the small magnitude events studied here do not resolve $Q$ properties at low frequencies smaller than \SI{3}{Hz},
our observations up to \SI{200}{Hz} exceed the high frequency limit of the other studies at least by a factor of two.
\par
The transport mean free path is the distance acrosswhich the propagation direction of $1{-}e^{-1}$ or \SI{63}{\%} of the wave energy becomes independent from its original propagation direction---the wave `forgets' its initial direction due to multiple scattering.
The inferred maximum values around \SI{300}{km} below 10~Hz decrease with frequency towards \SI{30}{km} at 200~Hz (Figure~\ref{fig:Q}a). 
Compared to the other studies the associated scattering strength $Q^{-1}\sca$ is relatively low (Figure~\ref{fig:Q}b), only the data collected in western India show consistently weaker scattering in the analyzed frequency range \citep{Ugalde2007}.
This means the average distance after which the wave propagation direction differs from the initial direction is comparatively large in the Fennoscandian Shield.
\par
The intrinsic absorption length is similarly defined as the length scale over which 63\% of the wave energy is dissipated.
One of the key results of this study is certainly the inferred long absorption length of around \SI{1000}{km} between \SI{8}{Hz} and \SI{30}{Hz} (Figure~\ref{fig:Q}a).
These very large values decrease by about two orders of magnitude to around \SI{30}{km} towards low frequencies and high frequencies. 
The exotic character of the results is highlighted in Figure~\ref{fig:Q}c, where the $Q^{-1}\intr$ values in the \SI{10}{Hz} to \SI{30}{Hz} range are an order of magnitude smaller than the lowest reported values from the reference cases.
This figure also shows an unparalleled decrease of $Q^{-1}\intr$ over 2.5 orders of magnitude between \SI{3}{Hz} and \SI{15}{Hz} that points to different relaxation processes at different scale lengths governed by the cratonic crustal structure and ambient crystalline rock properties.
The unusual frequency behavior of intrinsic absorption can directly be observed in the energy density envelopes (Figure~\ref{fig:data}b), where low and high frequencies show a fast coda decay, whereas the frequencies \SI{10}{Hz} to \SI{30}{Hz} show a slow coda decay.

\subsection{Site effects}
\label{sec:results_siteamp}

\begin{figure}
\centering
\includegraphics[width=\textwidth]{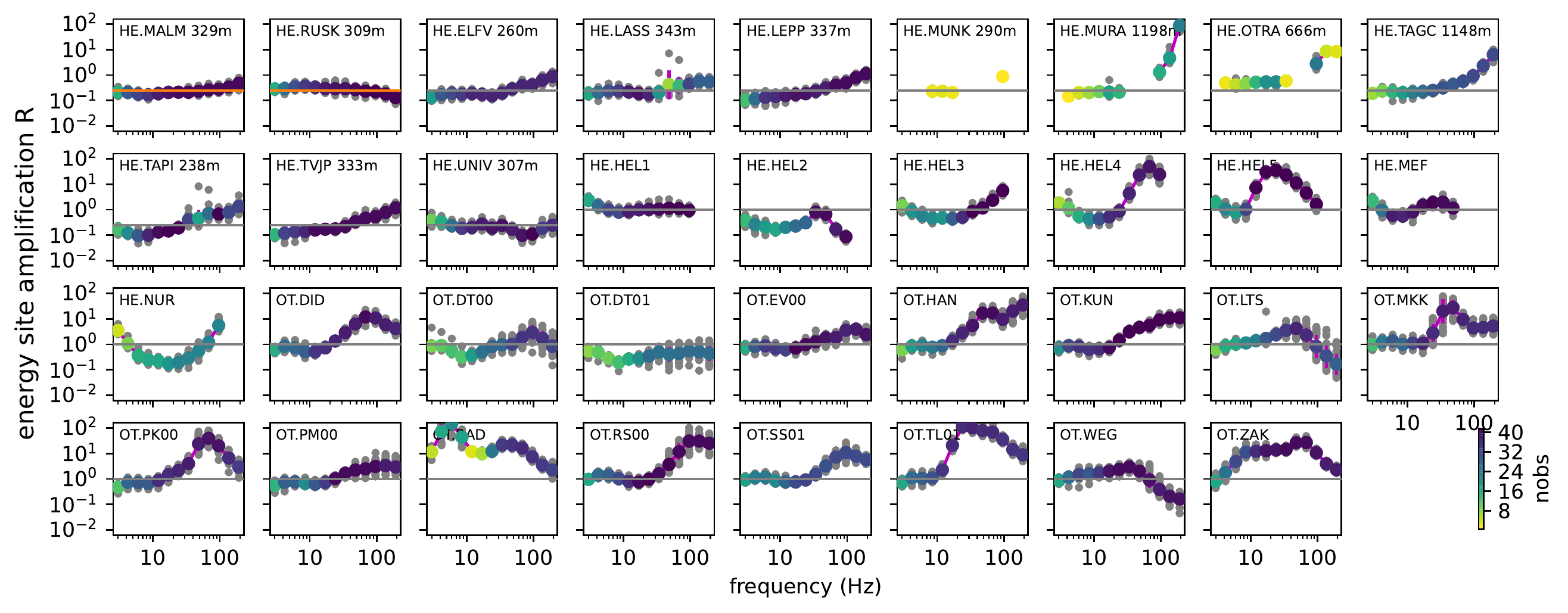}
\caption{Frequency dependent site terms $R(f)$ for the used stations. The first 12 panels MALM to UNIV show borehole station data, and the following 23 panels HEL1 to ZAK show surface station data. The sensor depth is indicated for the borehole stations. Grey dots indicate energy site amplification measured for a single earthquake. The geometric averages are indicated by colored circles. The color represents the number of earthquakes or observations nobs. The geometric mean of all observations of the two borehole stations MALM and RUSK is fixed to 0.25 in each frequency band which is indicated by the orange horizontal lines in the first two panels. Reference neutral site amplification values of 0.25 and 1 are indicated by horizontal grey lines for borehole stations and surface stations, respectively.}
\label{fig:sites}
\vspace{0.5cm}
\end{figure}

Figure~\ref{fig:sites} shows frequency dependent site terms. Panels are ordered left to right, top to bottom, and the first 12 panels MALM to UNIV show borehole station data. The terms of the borehole stations MALM and RUSK are fixed to an average of 0.25 for each frequency, again, as reference, to fix the trade-off between site and source effects. 
This 0.25 reference line is indicated in the borehole station panels.
It corresponds to a unit amplification considering the earlier applied free surface correction in Equation~\ref{eq:Eobs},
and the surface station panels indicate this reference unit value with a line, too.
Alternatives to this average frequency independent behavior and the effects on the obtained scaling relations are discussed in Appendix~\ref{sec:testRf}.
The site term values represent energy site amplification, i.e., the square of amplitude site amplification, compared to the reference level at the MALM and RUSK stations.
The grey indicated data in Figure \ref{fig:sites} are observations from individual earthquakes, colored data show averages.
The {\Ml}0.0 to {\Ml}1.8 magnitude range results in little variability around the average, which means the term `amplification' explicitly refers to the comparison with the reference. 
It does not include potentially nonlinear scaling effects with amplitude.
\par
The perfect trade-off between site and source terms means that a site term scaling at all stations by some factor results in an equivalent scaling of the spectral source energy by the inverse of that factor, and together they explain the observations just as well.
\citet{Eulenfeld2017} regionalized data from North America and observed a higher site amplification in the eastern part of the United States compared to the western part in the \SI{6}{Hz} to \SI{12}{Hz} range, which can result in biased moment magnitude estimates of small earthquakes if not taken into account. 
For the band-limited envelope of an event recorded at a given station, the envelope inversion using the global parameters $g$ and $b$ and the event-specific source term $W$ yields a fit to the data that cannot be improved by adjusting the site term $R$.
Correspondingly, for a given frequency, at a given site, a site term of $R{=}10$ indicates a 40-fold increase with respect to the 0.25 average at the two borehole reference stations and a 10-fold increase with respect to neutral amplification at surface stations.
The average of $R$ over all stations at any one frequency is not necessarily equal to the reference value.
Thus, the level of the $R(f)$ terms can vary depending on the choice of the reference, but the $R(f)$ shape at a given station is not sensitive to this choice as long as $R(f)$ fluctuations around unity at the reference sites are small.
\par
Borehole stations ELFV, LASS, and UNIV show a relatively flat site term at the 0.25 reference value that is not tuned at these locations.
Stations LEPP, TAPI, and TVJP equally show a flat site term at a level of 0.25 with a moderate increase at frequencies larger than about \SI{30}{Hz}.
Stations OTRA and TAGC show no relative amplification at low frequencies but increased values up to 10 towards higher frequencies.
Data at stations MURA and MUNK are overall of bad quality.
All sensors are located well below the low-velocity surface layer in a competent bedrock environment. 
Without more detailed knowledge about the deep site properties, we hypothesize that the high-frequency amplifications can be related to coupling effects.
\par
Relative amplifications are larger at surface stations, and they also show a more diverse behavior.
Surface stations HEL1, MEF, DT00, DT01, EV00, and PM00 show a relatively flat site term at the reference level over the whole frequency range.
Stations HEL3, DID, KUN, and SS01 show above-reference values at frequencies exceeding \SI{50}{Hz}.
Stations HEL4, HAN, and RS00 experience comparatively large terms up to 100 for high frequencies.
Station RAD shows elevated relative amplification across the full frequency range, especially at lower frequencies.
In contrast, station HEL2 exhibits below-reference site amplification at most frequencies.
The site amplification at station NUR shows a minimum of 0.2 around \SI{10}{Hz}, whereas
stations HEL4, HEL5, LTS, MKK, PK00, TLVJ, WEG, and ZAK show maxima in the amplification curves between \SI{20}{Hz} and \SI{80}{Hz}.
\par
The obtained data show that the relative spectral amplification is not uniform at most locations, and that larger effects are measured at surface stations.
As detailed in Section~\ref{sec:disc_siteamp}, these observations have implications for Fennoscandian ground motion prediction equations.
Since even the borehole acquisition in the 200~m depth range suggests frequency dependent site effects,
and since most surface stations are located on bedrock outcrops, the observations concern the practice of fixing the spectral amplification at Fennoscandian hard rock sites to unity \citep{Fueloep2020}.

\subsection{Displacement spectra and source parameters of induced earthquakes}
\label{sec:results_spectra}
\begin{figure}
\centering
\includegraphics[width=\textwidth]{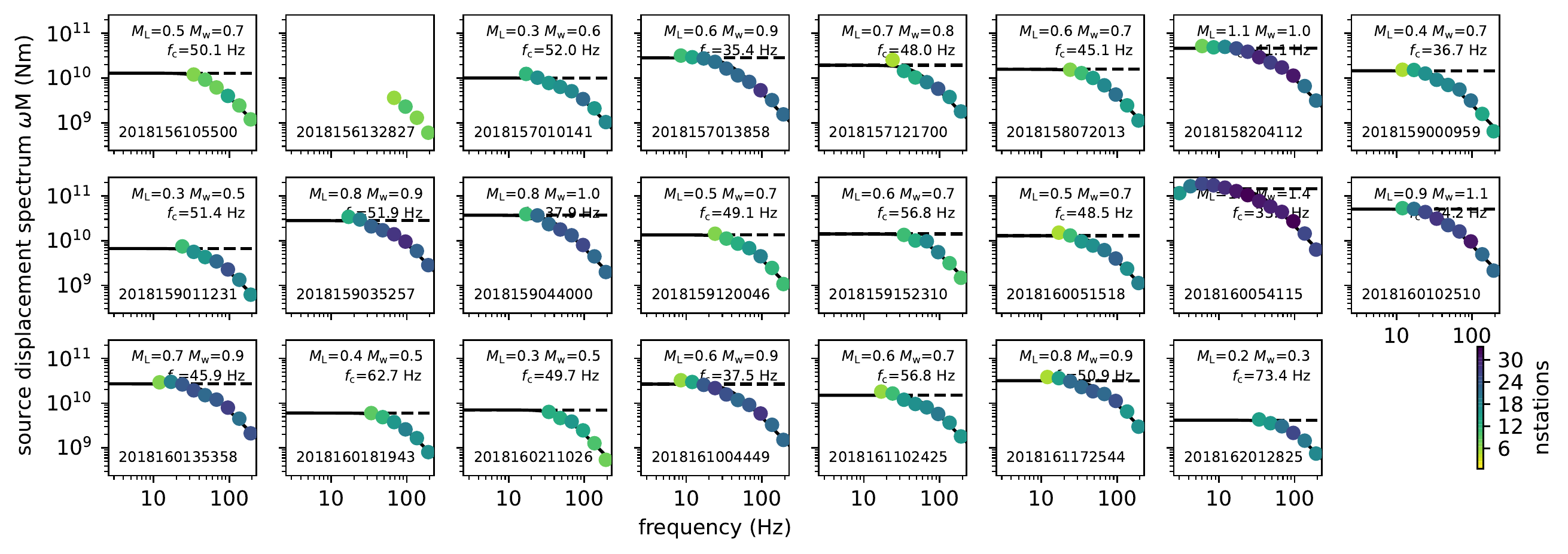}
\caption{Source displacement spectra $\omM\fof f$ of 23 selected earthquakes induced by the 2018 geothermal stimulation. Color indicates the number of used stations nstations. The seismic moments obtained by Qopen are shown by the dashed horizontal lines. Local magnitude, estimated moment magnitude, and corner frequency are indicated in each panel together with the earthquake identifier.}
\label{fig:sds}
\vspace{0.5cm}

\includegraphics[width=\textwidth]{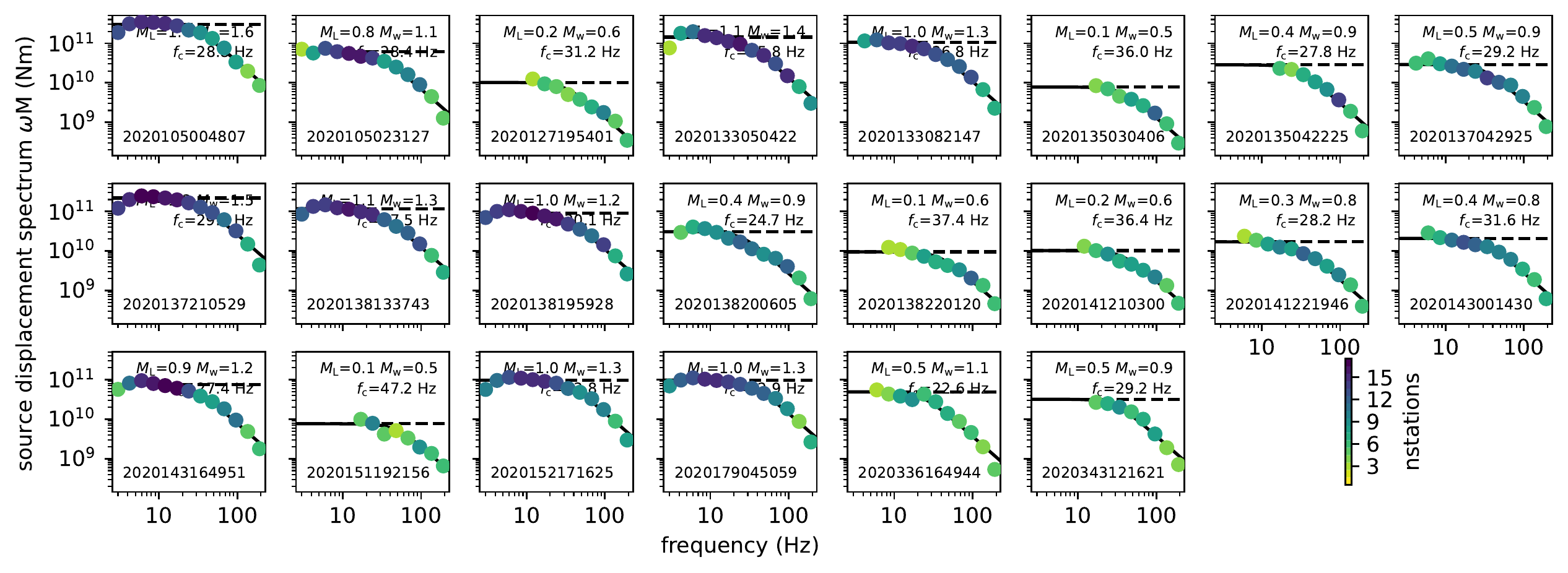}
\caption{Source displacement spectra $\omM\fof f$ for 22 analyzed earthquakes induced by the 2020 geothermal stimulation. The same conventions apply as in Figure~\ref{fig:sds}.}
\label{fig:sds2020}
\end{figure}

Figure~\ref{fig:sds} displays a selection of source displacement spectra obtained for earthquakes induced by the 2018 geothermal stimulation.
Figure~\ref{fig:sds2020} displays source displacement spectra of all analyzed earthquakes of the 2020 geothermal stimulation. Recall that the 2020 solutions were calculated using the Qopen monitoring mode, i.e., event locations and raw waveforms were used together with the estimates of attenuation parameters and site amplification factors obtained independently from the 2018 data.
Missing values at low frequencies are explained by the low signal amplitude at these frequencies for small events. 
The smoothness of the spectral shapes reflects the envelope, station, and frequency averaging (Equation~\ref{eq:Eobs}).
The similarity between spectra obtained from $S$~wave envelope analysis and direct $S$~wave spectra obtained from moment tensor analysis is demonstrated by the agreement of the derived source properties \citep{Eulenfeld2021}.
We note again that the Qopen Green's function approach results in physically accurate spectral amplitude values---provided the deployment facilitates a reasonable reference site term estimate---from which seismic moment and moment magnitude can be directly obtained, similar to the generalized inverse and EGF approaches, but different from the iterative stacking methods.
\par

Using the Boatwright-type model Equation~\ref{eq:sourcemodel} with $\gamma{=}2$ we estimate seismic moment $M_0$, moment magnitude \Mw, the high frequency falloff rate $n$, and the corner frequency $f\ind c$ from spectra of 209 (22) events induced by the 2018 (2020) stimulation that have more than four data points in $\omMf$. As detailed below we fix $n$ to its median value 1.74 for the final estimates of \fc, $M_0$ and \Mw. The stress drop $\Delta\sigma$ is estimated using the circular fault model from \citet{Madariaga1976} with $v\ind S{=}\SI{3.5}{km/s}$ in the hypocentral region
\begin{equation}
  \Delta\sigma = \frac 7{16}M_0\left(\frac{f\ind c}{kv\ind S}\right)^3\,\text{ with }k=0.21. \label{eq:stressdrop}
\end{equation}

\begin{figure}
\centering
\includegraphics[width=0.5\textwidth]{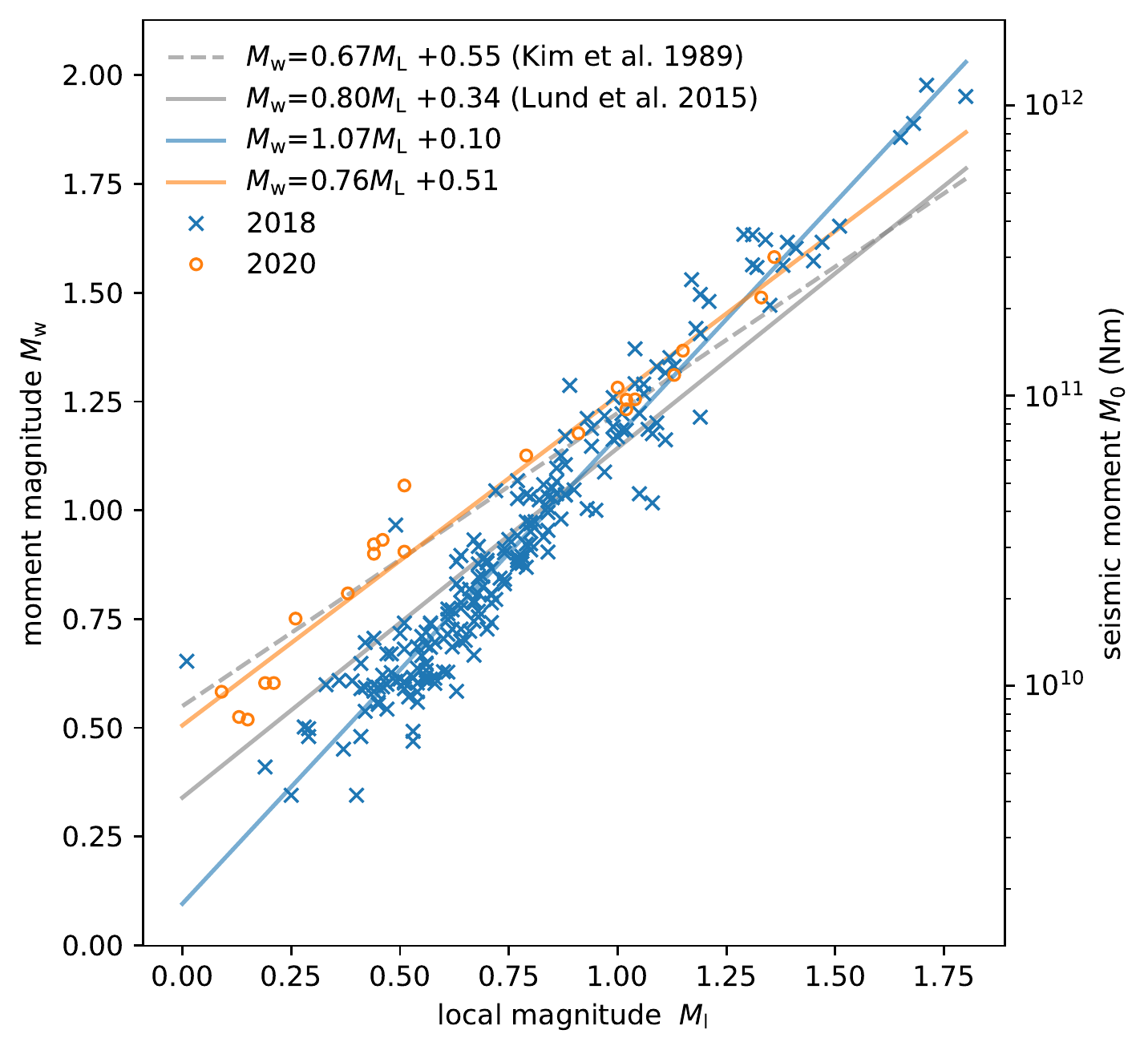}
\caption{Comparison of moment magnitude \Mw and local magnitude \Ml of the earthquakes from the 2018 and 2020 geothermal stimulation.
The least-squares fit to the 2018 and the 2020 data is indicated with a blue and orange line, respectively. The gray indicated relations determined in \citet{Kim1989} and \citet{Lund2015} are displayed for reference.}
\label{fig:mags}
\end{figure}

Figure~\ref{fig:mags} displays the scaling between moment magnitude \Mw obtained from $M_0$ \citep{Kanamori1977} and the associated local magnitude estimates \Ml for the earthquakes induced by the 2018 and 2020 stimulations. For the 2018 and 2020 data sets the \Mw-\Ml scaling is
\begin{eqnarray}
\Mw & = & 1.07 \Ml + 0.10 \\
\Mw & = & 0.76 \Ml + 0.51 \,.
\end{eqnarray}
In comparison, moment magnitudes for small events with $\Mw{<}1$ estimated from the 2020 data tend to be larger for the same local magnitudes.
Considering the consistent data processing of the 2018 and 2020 data, and the invariance of the $g$ and $b$ medium parameters obtained from the 2018 data that are equally applied to the 2020 data,
this difference suggests variations in the average source properties of the two earthquake populations.
For a perfectly elastic medium \Mw equals \Ml.
\citet{Deichmann2017} argues that for small earthquakes ($\Mw{<}2$ to 3) the \Mw-\Ml scaling is expected to be smaller than unity due to anelastic attenuation.
Therefore, our observed \Mw-\Ml slope $\sim$1 might be another manifestation of the extraordinary low intrinsic attenuation observed above \SI{7}{Hz}. The 0.76 proportionality factor of the 2020 scaling compares favorably with the 0.67 and 0.80 factors of the \Mw-\Ml relationships determined by \citet{Kim1989} and \citet{Lund2015} that are displayed for reference in Figure~\ref{fig:mags}.
These relations were also determined from data in the Fennoscandian Shield, albeit across larger areas and from a broader magnitude range.
\par

\begin{figure}
\centering
\includegraphics[width=1\textwidth]{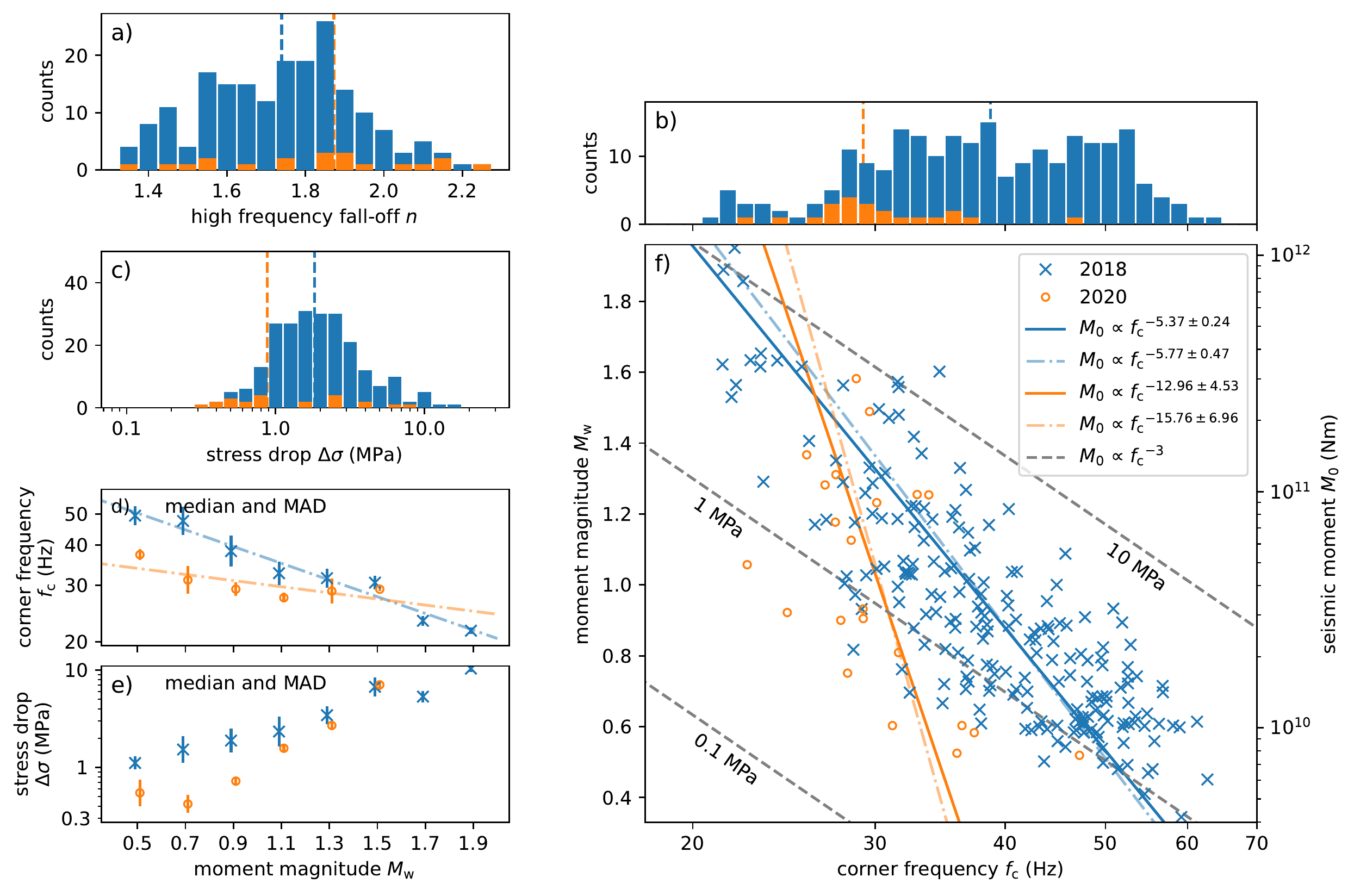}
\caption{Statistics of source parameters of 209 earthquakes induced by the 2018 stimulation indicated in blue and of 22 earthquakes induced by the 2020 stimulation indicated in orange. Displayed stress drops and corner frequencies are determined with a fixed high frequency falloff.
(a)~Histograms of observed high frequency falloff rates, (b) corner frequencies, and (c) stress drops of 2018 versus 2020 earthquakes. The vertical dashed lines in the histograms represent the medians of the distributions.
The mean and standard deviation for the 2018 data in (c) are \SI{2.5\pm4.5}{MPa}, and the median and the median absolute deviation around the median (MAD) is \SI{1.8\pm0.7}{MPa}. For the 2020 data, the mean and standard deviation are \SI{1.9\pm3.9}{MPa}, and the median and MAD are \SI{0.9\pm0.5}{MPa}.
(d)~Corner frequency and (e) stress drop estimated in different moment magnitude bins. The bins have a width of 0.2. Displayed are the median corner frequency and stress drop of 2018 and 2020 earthquakes in each magnitude bin. Error bars represent the MAD.
(f)~Scaling between moment magnitude and corner frequency.
Constant stress drops of \SI{0.1}{MPa}, \SI{1}{MPa}, and \SI{10}{MPa} corresponding to a scaling $M_0{\propto}\fc^{-3}$ are indicated with gray dashed lines \citep{Madariaga1976}. Continuous lines show the linear regression between moment magnitude and logarithmic corner frequency; dash-dotted lines in (d) and (f) show the linear regression between binned moment magnitude and the medians of logarithmic corner frequencies in the bins. Similar sized earthquakes induced by the 2020 geothermal stimulation appear to have a lower corner frequency and stress drop compared to the earthquakes induced by the 2018 stimulation.
}
\label{fig:fc}
\end{figure}

A systematic variation between 2018 and 2020 induced events is also implied by the other source scaling observations collected in  Figure~\ref{fig:fc}.
Figure~\ref{fig:fc}a shows the statistics of the inferred falloff rate $n$ values. For the 2018 population the median is 1.74, which is close to the classical omega-square model with $n{=}2$.
Because of the trade-off between \fc and $n$ \citep{Eulenfeld2016} we follow \citet{Eulenfeld2021} and fix $n{=}1.74$ for our final estimates of the corner frequency for the \Mw-\fc scaling and the inferred stress drop estimates (Figures~\ref{fig:fc}c--f).
As \citet{Kaneko2014} conclude from numerical tests, fixing $n$ can degrade the fit to individual spectra, but it helps to reduce bias in \fc associated with incomplete station coverage.
As mentioned, the high sampling rate supports the quality of the \fc estimates that are central and not near the edge or outside the fitting range \citep{Shearer2019,Abercrombie2021}.
\par
It has often been remarked that stress drop estimates are highly sensitive to the \fc value and the constant $k{=}0.21$ used in Equation~\ref{eq:stressdrop} that depends on a specific theoretical rupture model \citep{Dong2003,Cotton2013}, 
as the $f_c/kv\ind S$ ratio is cubed to estimate $\Delta\sigma$.
While this does complicate the appraisal of $\Delta\sigma$ scaling relations obtained in different studies making different assumptions, we think our here observed systematic variations are not controlled by ambiguities of this ratio.
In Figure~\ref{fig:fc}f we indicate lines of constant stress drop for a circular fault with a rupture velocity of 90\% of $v\ind S$ (Equation~\ref{eq:stressdrop}).
For constant stress drop the relationship between seismic moment and corner frequency is $M_0{\propto} \fc^{-3}$.
However, we observe a significantly higher decay rate of $-5.4$ and $-12.9$ for earthquakes induced by the 2018 and 2020 stimulation, respectively.
These steeper slopes indicate a higher stress drop for larger earthquakes compared to the extrapolation from small-event physics.
While the robustness of the slope estimate for 2020 events suffers from the limited amount of data, we consider the steep slope for the 2018 events significant because of the small standard error and because of the similar value obtained from a linear regression of binned data shown in Figure~\ref{fig:fc}d \citep{Supino2020}.
\par
Similar to the different \Mw-\Ml scaling relations in Figure~\ref{fig:mags} the different trends in the blue and orange data in the \Mw-\fc scaling Figure~\ref{fig:fc}f, too, suggest a systematic variation between 2018 and 2020 event properties.
Whereas the different \fc statistics in Figure~\ref{fig:fc}b alone are not indicative,
and while median stress drop values indicated in Figure~\ref{fig:fc}c are relatively similar for the 2018 and 2020 cases,
the data in Figures~\ref{fig:fc}d and \ref{fig:fc}e highlight the systematically lower corner frequencies and the correspondingly lower stress drop estimates for earthquakes with $\Mw{<}1$.
The mean stress drop and variance for the 2018 data associated with Figure~\ref{fig:fc}c is \SI{2.5\pm4.5}{MPa}. This 180\% variance is large, but the $\ln(4.5){=}1.50$ value is in the 1.4 to 1.7 range of values compiled by \citet{Cotton2013}.
The values for the 2020 data are \SI{1.9\pm3.9}{MPa}, a 200\% variance.
\par
To sum up, the \Mw-\Ml scaling (Figure~\ref{fig:mags}) shows different slopes for the 2018 and 2020 data, and both event populations exhibit \Mw-\fc relations (Figure~\ref{fig:fc}f) that are not compatible with size independent scaling.
The corner frequency dependence (Figure~\ref{fig:fc}d) translates to a stress drop dependence (Figure~\ref{fig:fc}e) on seismic moment.
In comparison to modern analyses of hundreds or thousands of earthquakes the combined sample size can be considered small, however, the inferred relations are sound within the available limits.
Systematic differences between 2018 and 2020 results in the \Ml, \fc, and $\Delta\sigma$ scaling with \Mw are discerned for $\Mw{<}1$. 
Whereas these trends can be considered the least robust due to the small 2020 event population, we include a discussion of potential driving mechanisms for these and the better confirmed average observations in the next section.

\section{Discussion}
\label{sec:disc}
\subsection{Crustal properties in the Fennoscandian Shield}
\label{sec:disc_crust}

The study area in southern Finland is located in the Uusimaa belt of the Paleoproterozoic Svecofennian domain, a part of the Fennoscandian Shield \citep{Lahtinen2012}.
The complicated internal structure of the Shield reflects the multistage Precambrian accretionary and orogenic deformation processes \citep{Lahtinen2005}.
The crust is thick, the Moho is about \SI{50}{km} deep \citep{Bruneton2004,Tiira2020}. 
The basement consists of igneous and metamorphic rocks of volcanic and sedimentary origin that were metamorphosed at \SI{1.9}{Ga} at \SI{15}{km} to \SI{20}{km} depth \citep{Pajunen2008}.
Sedimentary rock layers and several kilometers of crystalline rock have been eroded during Phanerozoic geological history.
The exhumed granites, gneisses, schists, and amphibolites are sheared and folded, they exhibit low porosity, the intergranular fluid content is very small, and fluids are mostly constrained to post-metamorphic brittle deformation structures \citep{Stober2007}.
Quaternary glaciations left a few meters thick layers of till and gravel, and layers of clay and peat were deposited during the Holocene in topographic depressions.
Tectonic movements, weathering, and deglaciation governed the post-orogenic brittle deformation that resulted in abundant lineaments, fractures, and faults.
The most prominent faults in the Helsinki area are the tens of kilometers long northeast-southwest-trending Porkkala-Mäntsälä fault and the north-south-trending Vuosaari-Korso fault \citep{Elminen2008}.
Elevation in the main deployment area (Figure~\ref{fig:map}a) varies between sea level and few tens of meters, with a horizontal wavelength in the \SIrange[range-phrase=--]{1}{5}{km} range.
For seismic wave propagation it is important that the subsurface is characterized by the hard, low-porosity crystalline rocks overlain by a patchy thin layer of soft sediments around frequent bedrock outcrops.
These features govern the observed comparatively weak high-frequency scattering and attenuation properties, and the variable site effects.
\subsubsection{Scattering and intrinsic attenuation properties}
\label{sec:disc_Qi}
Figure~\ref{fig:Q}a summarizes the observed partitioning of scattering and absorption effects using the respective scale length. The summary indicates that for frequencies below \SI{7}{Hz} intrinsic absorption dominates over scattering attenuation. In contrast, in the range between \SI{7}{Hz} and \SI{40}{Hz} scattering dominates over intrinsic attenuation with very large intrinsic attenuation length scale estimates of up to \SI{1000}{km}. 
For frequencies above \SI{50}{Hz} the contributions of intrinsic and scattering attenuation to the total attenuation are similar, with perhaps a slightly stronger effect of intrinsic attenuation.
\par
We emphasize that our scattering strength and intrinsic attenuation estimates for frequencies up to \SI{200}{Hz} extend the upper frequency limit of previous studies by several tens of Hertz.
To obtain robust Qopen results \citet{Laaten2021} suggest to use coda time windows with a minimum length of \SI{30}{s} based on their analysis of attenuation properties along the Leipzig–Regensburg fault zone in Germany. \Citet{Laaten2021} used envelopes of 18 earthquakes with local magnitude between 1.4 and 3.0 recorded at 20 broadband stations with event-station distances up to \SI{80}{km}. 
In comparison, our analysis here uses shorter event-station distances and our network convinces overall more with a better azimuthal station distribution (Figure~\ref{fig:map}). We are therefore confident that our shorter coda window length of \SI{18}{s} does not affect the quality of the estimates and the conclusions. 
We verified that an extension of the coda window to the limit which is allowed by the signal-to-noise ratio and that includes the diffuse Moho reflection between \SI{18}{s} and \SI{28}{s} (Figure~\ref{fig:data}a) does not vary our attenuation estimates significantly.
Coda envelope studies typically have to evaluate the effect of the Moho discontinuity because energy can leak into the mantle with a lower scattering coefficient or because energy can be trapped in the crust \citep{Margerin1998}, which can influence the $g$ and $b$ estimates. 
Here, the thick crust and the used short coda time window suggests an insensitivity to such potential Moho effects, and the appropriate application of the half-space assumption.
\par
The scattering attenuation $Q^{-1}\sca$ as a function of frequency (Figure~\ref{fig:Q}b) exhibits a typical shape that is also frequently observed in previous studies that use Qopen or a similar method. The frequency dependence can be parametrized with a power law $Q^{-1}\sca{\propto} f^{-2\nu}$ for frequencies above \SI{5}{Hz} \citep[page~176]{Sato2012}.
(We use $\nu$ instead of $\kappa$ to avoid confusion with another $\kappa$ parameter used below.)
Here, the medium heterogeneity can be described with a von~Kármán type random distribution, and the estimated $\nu{=}0.14$ quantifies the medium roughness \citep[chapter~2.3]{Sato2012}.
This is similar to the $\nu{=}0.11$ value obtained with very similar methods around the 9~km deep German Continental Deep Drilling project KTB \citep{Fielitz2015}.
Our observed transport mean free path limits of 300~km and 30~km at \SI{6}{Hz} and \SI{200}{Hz} are also comparable to the 340~km and 60~km values at \SI{6}{Hz} and \SI{72}{Hz} for the KTB environment.
Compositional heterogeneity obtained from 
$\sim$\SI{5}{km} deep borehole and vertical profile data at the boundary between the Svecofennian domain and the Transscandinavian igneous belt in Sweden has been interpreted using a spatial autocorrelation function (ACF) model with \SI{150}{m} correlation length, 1\% RMS (root mean square) velocity perturbation, and $\nu{=}0.29$ \citep{Line1998}.
\citet{Hock2000} applied the ACF approach to scattering results obtained from teleseismic $P$ coda signals to study lithosphere heterogeneity in the shield environment in southern Sweden across the Tornquist zone, estimating \SI{1}{km} correlation length and 4\% RMS velocity perturbations.
Their $Q^{-1}\sca$ values between 1/1000 to 1/300 in the \SIrange[range-phrase=--]{0.5}{7}{Hz} range are comparable to our $<$\SI{5}{Hz} low-frequency results but exceed our 1/3000 estimates at \SI{7}{Hz}.
The employed different models and obtained $Q^{-1}\sca$ values remind of the challenge to integrate scattering values and heterogeneity estimates from different locations obtained from different waveform features and frequency ranges, considering that variable resolution bands sample different windows of medium heterogeneity \citep{Line1998}.
\par
Intrinsic attenuation $Q\intr^{-1}$ values for frequencies below \SI{7}{Hz} are in the range of previously reported observations (Figure~\ref{fig:Q}c). 
In contrast, the unusually low $Q\intr^{-1}{<}\num{e-4}$ values between \SI{7}{Hz} and \SI{40}{Hz} imply that the probed crustal material of the cratonic Fennoscandian Shield converges towards perfect elasticity in this frequency range.
On a global scale, the second lowest attenuation observed in this band are the mean $Q\intr^{-1}$ values around \num{1.3e-4} at \SI{30}{Hz} for the oceanic crust in the Eastern North Atlantic \citep{Hannemann2021},
with $Q\intr^{-1}$ values approaching \num{0.9e-4} for the oldest sampled lithosphere.
For high frequencies above \SI{50}{Hz} intrinsic attenuation is again higher, but still overall low, with $Q\intr^{-1}{\approx} 10^{-4}$, but due to a lack of independent observations at this frequency range at other locations it is difficult to contextualize the values.
Intrinsic attenuation models involve relaxation mechanisms with characteristic time and hence frequency scales that are associated with characteristic rock element dimensions \citep[][chapter~5.2]{Sato2012}.
Although the low frequency results are less well constrained due to the smaller number of sufficiently large earthquakes, the here observed overall strong frequency dependence suggests a systematic change in the governing relaxation processes across the range of wavelength scales.
The significance of these results will benefit from complementary studies of other earthquake sequences in similar environments, e.g., in the Rapakivi granite area in southeastern Finland \citep{Luhta2022}, and from a comprehensive analysis of the seismicity across the Fennoscandian Shield \citep{Veikkolainen2021}.
Considerable depth dependent variations of $Q\intr^{-1}$ have been estimated using the borehole data from Sweden \citep{Line1998} using frequencies up to \SI{200}{Hz}.
Attenuation values at \SI{5}{Hz} along a 1981 deep seismic sounding profile in central Finland also vary between $Q\ind P{=}50{-}80$ and $Q\ind S{=}70{-}140$ in the topmost kilometer and $Q\ind P{=}80{-}800$ and $Q\ind S{=}140{-}300$ in a layer down to \SI{6}{km} depth \citep{Grad1994}.
This depth dependence is attributed to variable crack density, and since this likely also applies in the southern Finland study area, it can contribute to an explanation of the observed $Q\intr^{-1}$ frequency dependence.
However, the similarity of envelope shapes observed at borehole and surface stations (Figure~\ref{fig:fits}) implies that the here reported values reflect average properties in the sampled bulk of the medium, and are not artifacts governed by high-frequency trapping or guiding effects associated with the thin, shallow low-velocity layer.
\par
The comparatively low scattering strength and the factual absence of intrinsic attenuation in the Fennoscandian Shield 
facilitate deep imaging \citep{Line1998}.
These properties contribute to the high signal-to-noise ratio typically observed at seismic stations in Finland, where signals of small-magnitude events can be resolved at much greater distances compared to more dissipating environments in active plate boundary regions or in regions with sediment deposits.
The high transparency for seismic waves supports the high data quality of the International Monitoring System FINES station, and it facilitated one of the first observations of deep body waves reflections in short period ambient noise correlations \citep{Poli2012}.
\subsubsection{Diffuse reflections at crustal velocity contrasts}
\label{sec:disc_reflection}
We argue that the transient spectral energy increase between \SI{18}{s} and \SI{28}{s} after the origin time (Figure~\ref{fig:data}) is due to a diffuse reflection at the \SI{50}{km} deep Moho
that is also facilitated by the high seismic transparency,
although a direct Moho reflection is not observed.
The two-way Moho travel time for $P$~waves is ${\sim}\SI{15}{s}$, for $S$~waves ${\sim}\SI{25}{s}$.
The spectral energy decreases from the peak in the direct $S$~wave window to the level in the coda at $\sim$\SI{18}{s} after origin time by a factor of \num{e-4} (Figure~\ref{fig:data}).
The relative energy loss due to geometrical spreading of a direct wave that is totally reflected at the Moho is approximately \num{2.5e-3}, which is the squared ratio of a \SI{5}{km} event-station distance and the approximate \SI{100}{km} two-way distance to the Moho.
However, the reflection coefficient at the Moho is equal to or smaller than 2\% using velocity values below Finland, and for waves with steep incidence this yields a relative decrease in the envelope of a direct reflection that is at most \num{5e-5}, which is smaller than the observed relative reduction of the coda envelope by \num{e-4}. Therefore, direct waves reflected at the Moho cannot be observed in the ambient scattering regime that governs the observed coda level.
In contrast, scattered waves partly compensate the loss due to geometrical spreading, and Moho reflection coefficients can also be higher because scattering tends to increase the reflection angle at the Moho.  
Together these mechanisms can explain the observed transient energy increase as scattered energy reflected at the Moho.
\par
This interpretation is also compatible with the \SI{18}{s} and \SI{28}{s} timing and duration of the transient energy increase, since it is approximately limited by the two-way Moho travel time of ballistic $P$~waves and $S$~waves. 
The observation of this phenomenon benefits, again, from the overall low intrinsic attenuation.
The Moho reflection is not visible at high frequencies above \SI{100}{Hz} (Figure~\ref{fig:data}b) although the pre-event noise level is not yet reached at the time of its expected arrival. We think that the relatively shorter transport mean free path (Figure~\ref{fig:Q}) diffuses this signal at high frequencies.
The other, smaller increase in spectral energy density visible at \SI{10}{s} after the origin time 
can be attributed to a reflection at a \SI{20}{km} deep interface \citep{Tiira2020} that has been associated with the Conrad discontinuity \citep{Luosto1997}.
These observations might be an interesting target for future studies using Monte-Carlo simulations of scattered energy packets in a horizontally stratified medium \citep{Margerin1998, Lacombe2003}.

\subsubsection{Site effects}

\label{sec:disc_siteamp}
The long $S$~wave envelopes analyzed by the Qopen method allow a separation of source spectra, attenuation effects in the volume, and site effects associated with the local structure below a sensor.
Unaccounted for regional variations in site amplification can potentially bias moment magnitude estimates of small earthquakes \citep{Eulenfeld2017}.
The resolved site effect terms discussed in Figure~\ref{fig:sites} and Section~\ref{sec:results_siteamp} demonstrate that the spectral amplification relative to two chosen reference borehole sites is not neutral at the other borehole and surface sensors, that it varies as a function of frequency, and that the largest variations and the most diverse patterns are observed at surface stations and at frequencies larger than \SI{30}{Hz}.
\par
We iterate that the reference site is arbitrary, but the accuracy of the obtained amplification levels has been demonstrated by the similarity of Qopen moment estimates---which trade off with the amplification levels---with moment tensor derived moment estimates \citep{Eulenfeld2021}. 
\citet{Parolai2000} make a similar argument to support site term estimates from a reference method by comparing it to results obtained with the generalized inverse technique.
\par
Fennoscandian seismic hazard assessment and associated ground motion prediction equations are often motivated by seismic hazard analysis for nuclear facilities.
The focus has been on frequencies in the structurally hazardous range below \SI{10}{Hz}, and this limited range was partly controlled by data sampled at \SI{40}{Hz} or \SI{50}{Hz}.
This focus is now widened by an increased interest in geothermal energy production.
The data collected in this context can inform common practice \citep{Fueloep2020},
which assumes neutral spectral acceleration amplification at the locations of the broadband stations of the Finnish National Seismic Network or a regional network extension \citep{Kortstroem2018,Veikkolainen2021}.
These are typically quality installations at surface hard rock sites, which means all installations are considered to be of reference very hard rock site quality.
This relates to sites where the $S$ wave velocity in the upper \SI{30}{m} is \SI{2.8}{km/s} or larger.
Shallow bedrock seismic velocities in the Helsinki area fit this criterion \citep{Kortstroem2018,Tiira2020,Hillers2020},
but the $v\ind S$ reduction in the topmost $10{-}30$~m \citep{Hillers2020} imply that at least in the study area an average outcrop must not have necessarily have very hard rock site properties.
The resonance frequency of a vertically incident $S$~wave is approximated by $f_0{=}v\ind S / 4 h$ \citep[e.g.][]{Wegler1997}, which yields compatible relations between the \SIrange[range-phrase=--]{1}{2}{km/s} $v\ind S$ values in the topmost layer, the inferred layer thickness $h$, and the frequency range where the strongest site effects are observed.
\par
The stations HEL1 to HEL5 are also carefully installed broadband sensors, in contrast to the majority of the short period sensors that were placed in a sometimes only few tens of centimeters thin soil or peat layer on the outcropping rock sites in the Helsinki area \citep{Hillers2020,Rintamaeki2021}.
Together with the spatially variable properties of the imaged topmost low-velocity layer, this can perhaps partly influence the diversity of the obtained site effects. 
As said, the ground motion prediction equations ignore a site term. Even if we concede potentially spurious geophone coupling effects, 
the patterns resolved with the borehole sensors and the broadband instruments suggest that this conservative approach may be reconsidered in the future to explore the implications of separating source, medium, and site effects in the ground motion prediction equations.
This can extend the evaluation of shaking scenarios at high frequencies greater ${\sim}\SI{30}{Hz}$ where we observe the largest and most diverse amplification, although this frequency range may not be relevant for structural integrity.
A generally improved quantification of the attenuation and site amplification effects associated with wave propagation in the Fennoscandian Shield can also help explain better why even small magnitude earthquakes are observed at and reported from distances that are much larger compared to macroseismic observations collected in other tectonic environments \citep{Maentyniemi2017}.
\par
Our observations of site effects up to \SI{200}{Hz} relate to poorly understood processes governing high-frequency attenuation which is typically modeled using the $\kappa$ parameter in engineering seismology and ground motion studies \citep{Anderson1984,Ktenidou2014}.
This parameter can be separated into having a path and a site contribution, where the latter refers to effects associated with the shallow structure below a sensor. It is common to estimate a site-specific, zero-epicentral distance estimate, $\kappa_0$, for a study region.
It captures the region-specific effects of intrinsic attenuation and scattering attenuation in the shallow layer \citep{Parolai2015,Parolai2018},
and it is used in stochastic ground motion prediction equations.
Values of $\kappa$ or $\kappa_0$ are estimated from short direct $S$ waveforms or \SIrange[range-phrase=--]{1}{2}{s} long signals that include arrivals scattered in the topmost layer.
We see the potential that $Q\sca$ and $Q\intr$ obtained with Qopen using long envelope time windows, and which therefore represent the average properties of a crustal volume containing source and receiver locations, can constrain trade-offs between attenuation in the volume and local site effects \citep{Ktenidou2014} that include a combination of site amplification and near-surface attenuation \citep{Motazedian2006}.
We mentioned that the topmost low-velocity zone is not considered to affect the frequency dependent $Q$ shapes in Figure \ref{fig:Q},
but this layer can play a role in explaining the elevated site terms we derived.
Such effects can be further investigated using H/V spectral ratios
or spectral deconvolution, e.g., at the Elfvik site ${\sim}\SI{2}{km}$ to the northwest of the stimulation site, where the 250~m deep borehole sensor ELFV is located approximately below the EV array.

\subsection{Earthquake source parameter scaling relations} 
\label{sec:disc_scaling}
\subsubsection{Seismic moment - corner frequency}

We referred to the notion that induced and natural earthquakes are controlled by the same physics.
This entails the prevailing view that earthquakes are, on average, self-similar, a hypothesis that has also implications for hazard assessment and which is therefore relevant for induced seismicity studies.
Average self-similar scaling is suggested by numerous individual studies and by compilations of data obtained from various environments across a wide range of scales \citep{Abercrombie2021}.
Self-similarity implies that normalized and hence dimensionless lengths or velocities are constant, and that the final size of an earthquake cannot be inferred from properties during rupture initiation \citep{Aki1967}.
Self-similar relations of observed and simulated global source parameters include constant stress drop scaling and a seismic moment to corner frequency relation $M_0{\propto} f\ind c^{-3}$ \citep{Prieto2004,Ripperger2007}. 
\par
The here obtained exponents $-5.37{\pm}0.24$ and $-12.93{\pm}4.45$ for the 2018 and 2020 $M_0$-\fc scaling reflect that \fc increases with earthquake size relative to self-similar scaling (Figure~\ref{fig:fc}f).
We refer again to the small standard error and to the consistency of results obtained from a linear regression of binned data,
which together indicate that the general deviation from size-independent physics and the difference between the 2018 and 2020 scaling relations are significant.
This difference between the two stimulations is also seen in the systematic \fc offset between the corresponding populations in Figures~\ref{fig:fc}b, \ref{fig:fc}d, and \ref{fig:fc}f, and this, too, finds its equivalence in the stress drop scaling.
We first discuss possible trade-offs that could be responsible for an overinterpretation of the data, before we consider potentially relevant physical explanations.
 \par
A spurious masking effect associated with excess high-frequency attenuation would cause the observed shift of \fc towards smaller frequencies for smaller events in Figure~\ref{fig:fc}f.
Here, attenuation is not determined from spectra of the short direct pulse but independently from properties of the full waveform envelope.
\citet{Shearer2019} highlight that non-self-similarity can be a consequence of constraining the falloff rate $n$. This statement is made for empirical Green's function analyses, which has different sensitivities and trade-offs compared to the Qopen method.
However, the source model trade-off between $n$ and \fc applies here, too \citep{Eulenfeld2016}.
Recall that we fix $n{=}1.74$ for the 2018 and the 2020 source model fitting.
In Appendix~\ref{sec:testn} we discuss the trade-off between different choices of $n$ and the resulting \fc estimates (Figure~\ref{fig:testn}a), as well as the weak effect on the obtained $M_0$-\fc scaling relationship (Figure~\ref{fig:testn}b) and conclude that the $n{=}1.74$ choice does not control our results and conclusions.
\par
We discussed that the level of the frequency independent reference amplification
trades off with the spectral source energy and the resulting seismic moments, and that the choice of a different level does not affect the estimates of corner frequency. 
In Appendix~\ref{sec:testRf} we investigate the effects of
a frequency dependent reference site amplification on the fitted values of the falloff rate $n$, corner frequency \fc, stress drop $\Delta\sigma$, and consequently also on the $M_0$-\fc scaling (Figure~\ref{fig:testRf}) \citep{Trugman2017, Trugman2020}. 
We show that neither a linear nor an exponential reference site response model can make the associated $M_0$-\fc scaling relationships convincingly more compatible with self-similarity.
Here we continue to work with frequency independent reference site amplification terms of the 
approximately \SI{300}{m} deep borehole stations.
Other Qopen sensitivities that have not been thoroughly investigated but are beyond the scope of this work include the effect of the relatively sparse frequency sampling and the bandpass filtering that could be adopted from Gaussian filtering for surface wave dispersion analysis.
\par
A comprehensive revision of our results and an assessment of model choices, assumptions, and additional test results detailed in Appendix~\ref{sec:testn} and \ref{sec:testRf} together support the interpretation that physical mechanisms are relevant for the obtained non-self-similar earthquake scaling in the analyzed magnitude range.
Similarly steep \fc slopes have been reported for the 2000, 2008, and 2018 Bohemian earthquake swarms \citep{Michalek2013,Eulenfeld2021}.
All these observations are associated with fluid-driven sequences, which suggests that pore pressure or poro-elastic effects play a role, if we exclude significant bias associated with constraining the $n$ falloff (Appendix~\ref{sec:testn}).
The observed slope variation between the 2018 and 2020 data imply differences in the reservoir and thus event properties that can be related to the different stimulation dynamics.
Below we examine the consistency of the resulting non-self-similar stress drop scaling with the \Mw-\Ml scaling (Figure~\ref{fig:mags}) that differs significantly for 2018 and 2020 data, for which the $n$-\fc trade-off is irrelevant, and this consistency therefore suggests the inferred $M_0$-\fc trends are not governed by spurious effects.
\par
The $M_0$-\fc scaling plot of \citet[their Figure S6]{Kwiatek2019} collects 56 data points associated with {\Mw}0.9 to {\Mw}1.9 events induced during the 2018 stimulation.
Visual inspection suggests a better agreement of these results with the constant stress drop line and hence a better compatibility with the self-similarity hypothesis compared to our results.
This disagreement is likely associated with different processing strategies.
The \citet{Kwiatek2019} analysis of ${\sim}\SI{0.5}{s}$ long $P$ and $S$ wave arrivals to estimate displacement source spectra and the applied parameter estimation differs from the Qopen approach.
\citet{Kwiatek2019} work with frequency independent attenuation.
Our results obtained with the envelope analysis demonstrate that the assumption of frequency-independent attenuation does not hold, and even a limitation to $f{>}10$~Hz may take the observed $Q(f)$ variations into account.
We demonstrate in Appendix~\ref{sec:testRf} how systematic site-term related effects change the source spectra shapes and hence the $M_0$-\fc scaling. We deduce that $Q$-related shape changes can similarly influence the trade-off in the joint inversion for moment, corner frequency, and attenuation in \citet{Kwiatek2019}.
This disagreement highlights the challenge to achieve convergence of results obtained from the same events using different signals and inversion techniques, and to comprehensively assess the assumptions, sensitivities, and biases of the involved methods.

\subsubsection{Seismic moment - stress drop}

We continue with stress drop scaling and a comparison to the diverse stress drop variation studies. Stress drop $\Delta\sigma$, moment $M_0$, and corner frequency \fc are related through $\Delta\sigma{\propto} M_0\fof{\fc/k v\ind S}^3$ (Equation~\ref{eq:stressdrop}),
which entails different effects of the errors in \fc and $M_0$ on the error of the $\Delta\sigma$ estimate \citep{Cotton2013}. 
In the previous sections we detailed the deviation in the $M_0$-\fc scaling from self-similarity for both stimulations,
and the consistent difference of the scaling exponents and the offset along the \fc axis for the 2018 and the 2020 events.
Figure \ref{fig:fc}e shows three stress drop scaling trends that correspond to these \fc patterns.
First, both the blue 2018 and the orange 2020 data show an increase in stress drop with moment magnitude in the \Mw range 0.5--1.9.
The \fc scaling trend associated with the steeper $M_0$-\fc slope (Figure \ref{fig:fc}d) thus translates to the observation that larger earthquakes exhibit a proportionally larger stress drop compared to smaller earthquakes.
Second, the slopes of the $\Delta\sigma$-$M_0$ 2018 and 2020 data differ, which is related to the different \fc scaling exponents (Figure \ref{fig:fc}f).
Third, the 2020 stress drop estimates are systematically lower compared to the 2018 data, at least for $\Mw{<}1$, which is related to the corresponding offset along the \fc axis (Figure \ref{fig:fc}f).
\par
The observed \Mw-dependent stress drop scaling for the 2018 and the 2020 data has practical implications since stress drop is an input parameter for seismic hazard studies, and it is often assumed to follow self-similar scaling \citep{Graves2010,Aagaard2010,Cotton2013}.
Hence although consolidating observations suggest moment release to depend on the injected volume during hydraulic stimulation \citep{Bentz2020,Kwiatek2022}, deviations from self-similarity would require magnitude dependent shaking limits that are adapted to the scale dependent radiation.
The resulting assessments can impact the evolving legislation and regulation of geothermal energy production associated with an increase in anthropogenic seismic activity, in particular in a low-background seismicity environment such as the Fennoscandian Shield.
As discussed in the previous section, we consider that the inferred deviation from self-similarity can be influenced by fluid related mechanisms.
The effect of fluids can further play a role in the 
variability of the $M_0$-$\Delta\sigma$ scaling, i.e., the different slopes inferred from the blue 2018 and orange 2020 data in Figure~\ref{fig:fc}e, because of the connection to the different injection schemes.
The observed variability, however, can be influenced by the small sample size of the 2020 data, similar to the observed difference between the 2018 and 2020 stress drop estimates at small magnitudes.
We acknowledge the controversial character of these two features and hesitate to extrapolate these trends to larger events beyond the here analyzed frequency range, or to earthquake behavior in general.
However, we recognize the significance of these variations and trends in relation to the formal uncertainties,
and continue with a discussion of the implications and potential driving mechanisms.
\par
Numerous studies discuss stress drop variations, but a single, consistent, universal controlling mechanism has not been established, which more likely reflects the complexity of earthquake faulting than complexities associated with data acquisition and quality, processing choices, and model assumptions.
Candidates for controlling mechanisms include earthquake size and mechanism, depth, tectonic setting, and fluid related properties.
The stimulated Otaniemi reservoir has been interpreted as a relatively homogeneous fracture network \citep{Kwiatek2019,Kwiatek2022}. Even so the reservoir is permeated by damage zones that govern the fluid flow and hence affect the location of the three main clusters in relation to the open hole sections for the 2018 stimulation \citep{Kwiatek2019}.
The proximity between the 2018 and the 2020 seismicity clusters highlighted in Section~\ref{sec:intro} implies that systematic variations in the structural reservoir properties do not govern the observed stress drop scaling differences. 
Instead, within the framework of the here adopted set of equations we discuss two physically consistent mechanisms that link the stress drop variation to fluid effects, and that are compatible with the observed trends.
These mechanisms are systematically different slip speeds and a reduced shear modulus.
\par

\begin{figure}
\centering
\includegraphics[width=0.7\textwidth]{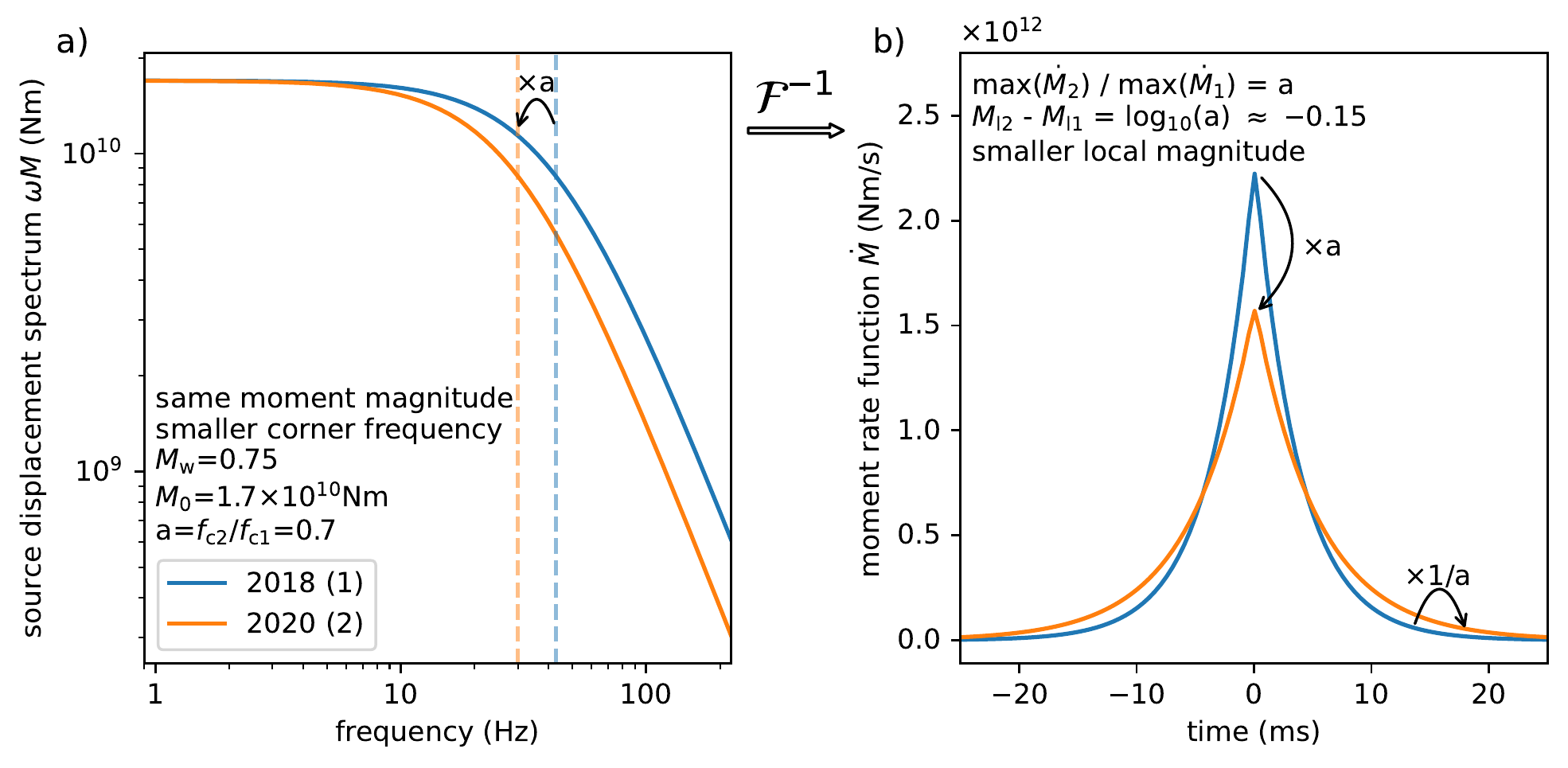}
\caption{Two earthquakes with the same seismic moment but different corner frequencies correspondingly have a different local magnitude.
(a) Source displacement spectra for two example earthquakes with moment magnitude $\Mw{=}0.75$ and corner frequencies $f_{\text c1}{=}\SI{42.9}{Hz}$ and $f_{\text c2}{=}\SI{30.0}{Hz}$.
Indices 1 and 2 correspond to blue 2018 and orange 2020 example data.
Corner frequencies are indicated by dashed lines. The ratio of corner frequencies is $a{=}0.7$. Seismic moment $M_0$ is proportional to the low frequency plateau.
(b) The associated moment rate functions are calculated by the inverse Fourier transform $\mathcal F^{-1}$ assuming constant phase for all frequencies. The seismic moment is 
$M_0{=}\int_{-\infty}^\infty \dot M_1\text dt{=} \int_{-\infty}^\infty \dot M_2\text dt$.
The moment rate function of the 2020 event is reduced by a factor $a$ and stretched in time---it is slower---by a factor $1/a$ compared to the moment rate function of the 2018 event.
The difference in local magnitude between both events is $\log_{10}a{\approx} -0.15$ due to the different amplitudes of moment rate functions.
}
\label{fig:fourier}
\end{figure}

First we argue that the smaller 2020 stress drop estimates compared to the 2018 estimates for small magnitudes in Figure~\ref{fig:fc}e are consistent with the different \Mw-\Ml scaling in Figure~\ref{fig:mags} and the interpretation that the 2020 events feature slower slip speeds.
The lower corner frequency and stress drop of earthquakes induced by the 2020 stimulation compared to earthquakes with the same moment magnitude induced by the 2018 stimulation corresponds to the observed smaller local magnitude of 2020 earthquakes compared to similar moment-sized 2018 earthquakes (the orange line is shifted to the left of the blue line for $\Mw{<}1$ in Figure~\ref{fig:mags}). 
Consider a ratio of corner frequencies of 2020 (index 2) versus 2018 (index 1) earthquakes $a{=}f\ind{c2}/f\ind{c1}{\approx} 0.7$ as inferred from the regression lines for {\Mw}0.75 events (Figure~\ref{fig:fc}d). For this argument we assume that the source displacement spectra are similar and can be described by a stretching of the frequency axis, i.e., $\omM_2\fof f{=}\omM_1\fof{f/a}$ (Figure~\ref{fig:fourier}a).

The inverse Fourier transform of the source displacement spectrum is the moment rate function $\dot M\fof t$ (Figure~\ref{fig:fourier}b). From the scaling relationship of the Fourier transform follows $\dot M_2\fof t{=}a\dot M_1\fof{at}$, i.e., the moment rate functions of 2020 events are a stretched version of the moment rate functions of 2018 events---2020 events release the same moment over a longer time, they have slower slip speeds---with a peak amplitude decrease of a factor $a$ for similar sized earthquakes (Figure~\ref{fig:fourier}). Local magnitudes are determined on a logarithmic scale from the maxima of displacement which are proportional to the amplitude of the moment rate functions \citep[e.g.][page~186]{Sato2012}. The expected difference in local magnitude is therefore $\log_{10}a\,{\approx}-0.15$ which is of the same order as the observed difference in local magnitude of 2020 versus 2018 similar sized earthquakes (Figure~\ref{fig:mags}).
\par
This discussion establishes that the observed small-\Mw trends in the \Mw-\Ml, $\Delta\sigma$-\Mw, and \Mw-\fc scaling (Figures \ref{fig:mags}, \ref{fig:fc}d, \ref{fig:fc}e, \ref{fig:fc}f) are compatible with slower slip speeds for the 2020 low-stress drop events.
Why do the 2020 events, on average, exhibit a more sluggish behavior?
As said, the 2020 stimulation was less energetic on average. 
Peak well-head pressure was 20\% smaller, and only 15\% of the 2018 water volume was injected, which plausibly translates to smaller pore pressures in the underground reservoir. 
For the Helsinki stimulations, high and low pore pressure and stress drop thus correlate.
This is opposite to examples from the literature
where we consider the reported distance-pressure pattern analogous to our time-pressure dependence to evaluate the pressure-stress drop scaling.
During EGS stimulations in Basel, Switzerland, and near Dallas, U.S., for a carbon capture and storage project near Decatur, U.S., and for the Berlín geothermal field, El Salvador, low stress drop values of events near the injection point correlate with high pore pressure perturbations in these regions,
and stress drop increases away from the injection as the pore pressure decreases
\citep{GoertzAllmann2011,Kwiatek2014,GoertzAllmann2017,Jeong2022}.
For a stimulation in Soultz-sous-For\^ets, France, near-repeating events exhibit scaling relations 
that suggest lower stress drops are similarly associated with higher fluid pressures \citep{Lengline2014}.
In all cases, a reduction in normal or differential stress through elevated pore pressure was concluded to control the reduced stress drop observations which is the inverse of our observed dependencies.
However, since we focus here on average trends---and this refers to the 2018 to 2020 stress drop difference and to stimulation parameters---we cannot rule out that a more detailed event-by-event analysis in relation to better resolved local ambient conditions can potentially resolve these inconsistent observations.
An alternative explanation for the more sluggish behavior of the 2020 events includes that the hypothesized slower slip speeds are controlled by fluid assisted stabilization of induced events,
a mechanism that has been discussed for properties of slip events in subduction zone environments \citep{Lengline2014}.
This scenario considers the water volume injected during the 2018 stimulation.
The 2020 seismicity is located at the edge of the 2018 seismicity \citep{Leonhardt2020} and hence at the edge of or behind the fluid diffusion front, which implies a potential influence of the excess water content and the associated changes in rock properties on the estimated source properties.
\par
A second mechanism that is compatible with the lower 2020 stress drop values also considers excess fluid effects.
Equation~\ref{eq:stressdrop} implies that a reduced shear modulus and hence $S$~wave velocity $v\ind S$ in the source region of the 2020 events could equally well explain our observations, because the rupture velocity directly depends on the seismic velocity in the applied rupture model of \citet{Madariaga1976}. 
In this case, a relative change in corner frequencies with factor $a$ can be translated to a relative change in seismic velocity with the same factor $a$. In contrast, the same change in corner frequency can only be explained by a systematic stress drop decrease by a factor of $a^3$. For $a{=}0.7$ this yields values of 70\% of the original velocity and 34\% of the original stress drop.
Related to this connection between $v\ind S$ and rupture velocity is the choice of the $k$ parameter in Equation~\ref{eq:stressdrop} \citep{Cotton2013}.
If fluid related changes in rock properties or the local stress field lead to consistent, systematic changes in the rupture speed or rupture mode, the constant $k{=}0.21$ value \citep{Dong2003} and hence the stress drop estimates had to be revised. 
Depending on the effect on $k$, this entails the opposite scenarios that the difference in the stress drop levels in 2018 and 2020 is either spurious or underestimated.
Another explanation involves unaccounted for systematic medium changes in the source region related to attenuation that the coda waves are not sensitive to, and that are potentially folded into the displacement source spectra.
\par
Finally, we think we can rule out that the obtained variations are biased by changes in the acquisition. 
The majority of the stations shown in Figure~\ref{fig:map}a was used for the 2018 and 2020 data analysis.
For both cases the network is complete enough to suggest that systematic biases associated with poor azimuthal averaging or generally a too small number of stations do not govern the observed trends and patterns
\citep{Kaneko2014,Shearer2019},
even if the events exhibit asymmetric ruptures \citep{Kaneko2015,Holmgren2023}.
\par
While distinctively different properties of human induced and natural seismicity on the event level have not been resolved,
statistical features reveal systematic differences of inter-event relations in stimulated and in natural earthquake clusters \citep{Zaliapin2016}.
Analysis of cluster features has been first applied to highlight the connection of different statistics to variable physical properties of the crust at regional scales \citep{Zaliapin2008,Zaliapin2013}.
Cluster analysis of seismicity induced during both stimulations suggests only limited earthquake interaction and hence the activation of an existing fracture network \citep{Kwiatek2019,Kwiatek2022}, in contrast to high event interaction and triggering during fresh fracture creation.
The overall similar responses of the 2018 and the 2020 stimulations in terms of the inferred structural reservoir properties indicate again that the variations observed here are not associated with systematically different fracture distributions.
Fluid volumes have been shown to influence the type of clustering \citep{Zaliapin2013}.
These approaches can be combined with observations of event or cluster behavior such as the here discussed stress drop variations to test hypotheses about energy release patterns during different stimulation stages and their relation to evolving reservoir properties \citep{Kwiatek2019}.
\par
In summary, the obtained scaling relations exhibit a systematic difference between the 2018 and the 2020 results---something did change.
Assuming that spectral fitting trade-offs and sample size effects do not govern the observations,
we demonstrated that fluid related different slip speeds, systematic changes in rock properties, or a combination of both are plausible physical scenarios that are compatible with the average stress drop behavior. 
Even if the systematic stress drop variations are biased by the assumption of similar rupture models, the need to adjust these assumptions or to update the medium parameters for the 2020 stimulations highlights an interesting evolution of the response.
To further constrain these results and associated trade-offs obtained with the Qopen approach, complementary source spectra estimates and their alternative processing for parameter estimation can be applied to data from the two stimulations.
A more extended analysis can target the resolution of space-time patterns within each sequence and the relation of such patterns to the, first, time variable pumping parameters logged in the engine room, and to the, second, modeled spatio-temporal evolution of pore pressure and poro-elastic effects in the stimulated volume.
Lapse-time tomography \citep{Calo2013} or passive imaging \citep{Hillers2015a} to resolve seismic velocity variations indicative of elastic material changes and fluid saturation
can further reduce the trade-off between source and medium effects.

\section{Conclusions}

The Qopen method \citep{SensSchoenfelder2006a,Eulenfeld2016,Eulenfeld2021,Qopen} provides an internally consistent radiative transfer based Green's function modeling approach to estimate earthquake source spectra, average medium attenuation parameters, and site effect terms from waveform envelopes. Here the application to seismograms from {\Ml}0.0 to {\Ml}1.8 earthquakes induced by two geothermal stimulation experiments in southern Finland \citep{Kwiatek2019,Hillers2020,Leonhardt2020,Rintamaeki2021} yields an array of diverse results in the \SI{3}{Hz} to \SI{200}{Hz} range with implications for source studies, hazard assessment and ground motion modeling, wave propagation in cratonic shield areas, and imaging and monitoring.
We established the utility of the Qopen approach as real-time monitoring tool \citep{Eulenfeld2021}.
\par
The Qopen approach isolates displacement source spectra differently compared to established methods such as the generalized inverse technique \citep{Andrews1986,Parolai2000}, iterative stacking schemes \citep{Prieto2004,Shearer2006,Trugman2020}, or the empirical Green's function method \citep{Berckhemer1962,Mueller1985,Ruhl2017}. Obtaining attenuation from long envelope signals is regarded advantageous for robust source spectra estimates. The complementary source parameter estimates provided by the Qopen technique can help reconcile apparent or genuine source scaling patterns obtained in different environments with different techniques.
\par
Results from the 2018 and 2020 Helsinki stimulations suggest systematic differences in the corner frequency, stress drop, and magnitude scaling relationships. 
Together with the consistently observed deviation from self-similar scaling that is possibly linked to fluid-related mechanisms, and the frequency and station dependent site effect terms, these results highlight the need for region-specific hazard assessment and ground motion modeling related to hydraulic stimulations.
\par
The induced waveform coda analysis shows that intrinsic absorption is exceptionally low in the crystalline low-porosity hard bedrock environment of southern Finland, where $Q\intr^{-1}$ values as small as \num{2e-5} at \SI{20}{Hz} imply near-perfect elasticity.
This has been facilitating crustal imaging in the Fennoscandian Shield \citep{Grad1994,Line1998,Poli2012,Tiira2020}, and here it helps explain the observation of a transient signal associated with the diffuse reflection at the ${\sim}\SI{50}{km}$ deep Moho discontinuity.
\par
The associated high signal quality is anticipated to support lapse time earthquake tomography \citep{Calo2013} to image the evolving reservoir properties,
and coda waveform based inter-source interferometry \citep{Snieder2005,Eulenfeld2020}
to help constrain the relative contributions of local medium changes and earthquake source effects on the obtained scaling relations.
Noise based monitoring and imaging can also provide independent observations for improved trade-off mitigation \citep{Hillers2015a,Obermann2015}.
The application of these passive imaging methods will benefit from the obtained good estimates of the scattering properties and the overall crustal structure for the construction of region specific scattered wave propagation models \citep{Paasschens1997,Kanu2015,Obermann2016}.
Together, the compact source region, the relatively homogeneous medium, the quality network, and the application of a diverse range of analysis techniques using ballistic and scattered wavefields have the potential to better constrain feedback mechanisms between engineered subsurface changes and induced event properties.

\label{sec:conc}

\appendix

\section{Trade-off between high-frequency falloff, corner frequency, and source scaling exponent}
\label{sec:testn}
\begin{figure}
\includegraphics[width=\textwidth]{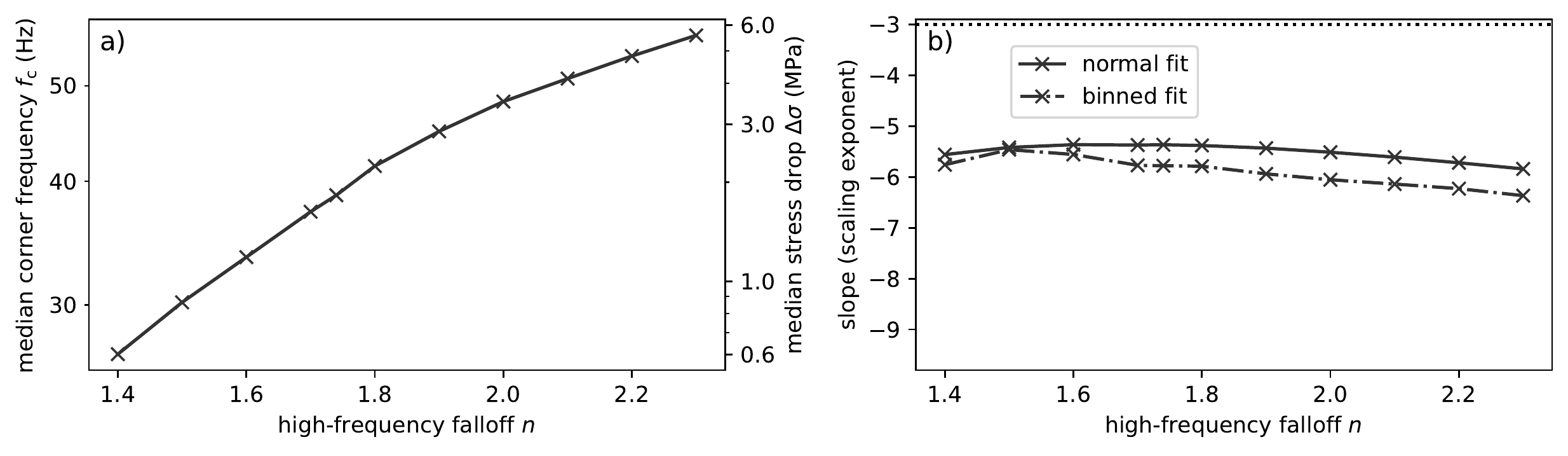}
\caption{(a) Trade-off between high-frequency falloff $n$ and median corner frequency \fc and median stress drop $\Delta\sigma$. The median refers to the median value from the 209 inverted events that occurred in 2018. The stress drop is calculated from the corner frequency assuming a moment magnitude {\Mw}0.9. (b) Trade-off between $n$ and the slope of the scaling relation between the corner frequency and seismic moment (compare to Figure~\ref{fig:fc}f). The slope of the linear regression (binned regression) that corresponds to the blue continuous (dash-dotted) line in Figure~\ref{fig:fc}f is indicated by the continuous (dash-dotted) line. A slope value 
of $-3$ associated with self-similarity is indicated by the dotted line for reference.
}
\label{fig:testn}
\end{figure}

In our source spectra analysis (Section~\ref{sec:results_spectra}) we first invert the obtained spectra for source parameters including the high-frequency falloff $n$. In a second step, $n$ is fixed to the median $n{=}1.74$ to avoid the trade-off between $n$ and the corner frequency \fc. Here, we test the impact of different choices of $n$ on the \fc estimates, on the associated stress drop $\Delta\sigma$, and on the $M_0$-\fc scaling (Figure~\ref{fig:testn}). There is a strong correlation between $n$ and \fc (Figure~\ref{fig:testn}a), a less steep high-frequency falloff corresponds to lower corner frequencies and smaller stress drop estimates, and vice versa.
More importantly, however, the choice of $n$ does not significantly influence the scaling between moment magnitude and corner frequency 
(Figure~\ref{fig:testn}b). 
All obtained $M_0$-\fc scaling slope values are in the $-5$ to $-6$ range and thus differ significantly from the $-3$ value associated with self-similarity.
The slope values obtained from the fits to the 209 data points do not change significantly from the $-5.37$ value of the original analysis (Figure~\ref{fig:fc}) because the same value of $n$ is applied to each spectrum, i.e., it systematically affects all measurements.
This observation suggests that our conclusion on the deviation from self-similarity is not controlled by the choice of the falloff rate $n{=}1.74$.

\section{The effect of a frequency dependent site amplification on high-frequency falloff, corner frequency, and source scaling exponent}
\label{sec:testRf}

In our main analysis, we fix the geometric mean of the site amplifications of the reference borehole stations MALM and RUSK to 0.25 independent of 
frequency (Section~\ref{sec:results_siteamp}). 
The associated flat reference site terms $R(f)$ for MALM and RUSK are shown in Figure~\ref{fig:sites}.
The 0.25 value corresponds to neutral unit amplification with the applied surface correction. 
Frequency independence is a plausible working hypothesis considering the borehole stations are located at a depth of around \SI{300}{m} in bedrock.
Here, we test the impact of a potential frequency dependence of the reference amplification. 
Our linear and exponential test functions approximate trends in the observed site terms, e.g., at stations LEPP, MURA, TAGC, TAPI, TVJP (Figure~\ref{fig:sites}).
These observations caution to take such effects into consideration, which can potentially also include narrow-band resonance effects that are, however, not evaluated here.
We test the impact on the resulting site terms $R(f)$ that consequently affect the spectral source estimates $W(f)$ since only the product $R(f)W(f)$ is constrained by the data.
In turn, systematic changes in $W(f)$ can affect the source parameter scaling relations that we observe and interpret to deviate from self-similarity.
\par
\begin{figure}
\includegraphics[width=\textwidth]{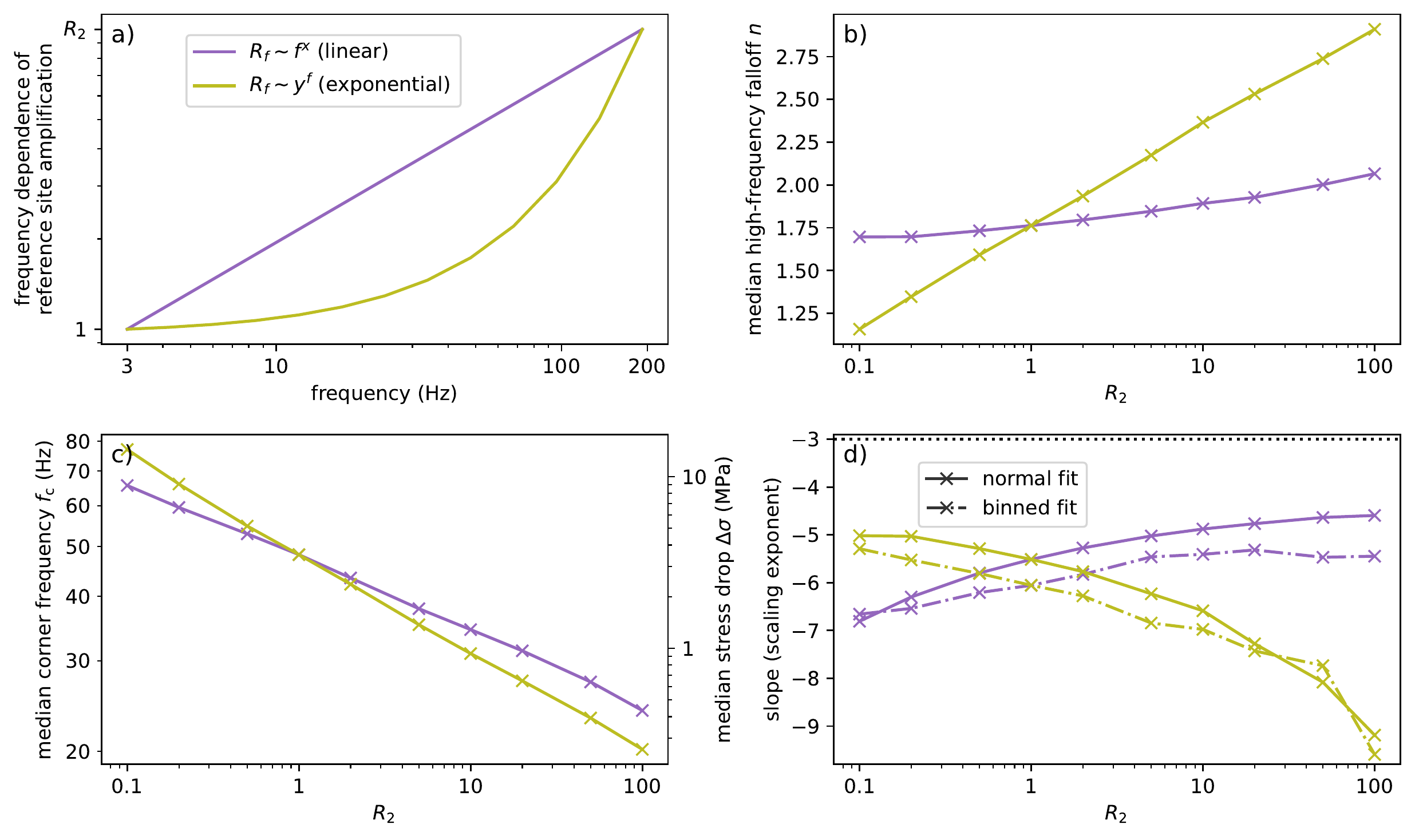}
\caption{(a) Reference site amplification models with a frequency dependence that follows a power law. The frequency is in the base (purple) or exponent 
(green) of the power law, $x$ and $y$ are constants. The same convention applies to the other panels. The maximum or minimum energy site amplification $R_2$ at \SI{192}{Hz} varies between 0.1 and 100.
(b) Trade-off between $R_2$ and the high-frequency falloff $n$.
(c) Trade-off between $R_2$ and the median corner frequency \fc as well as the median stress drop $\Delta\sigma$. The stress drop is calculated from \fc assuming a moment magnitude {\Mw}0.9.
(d) Trade-off between $R_2$ and the slope of the $M_0$-\fc scaling relation. The slope of the standard linear regression (binned regression) that corresponds to the blue continuous (dash-dotted) line in Figure~\ref{fig:fc}f is here indicated by the continuous (dash-dotted) lines. The slope of $-3$ associated with self-similarity is indicated by the dotted line for reference. In the analysis shown in panels (c) and (d) the high frequency falloff is fixed to $n{=}2$.
}
\label{fig:testRf}
\end{figure}

We introduce an additional frequency dependent site amplification factor $R_f$. 
For this test we choose $R_f\fof{\SI{3}{Hz}}{=}1$ and vary $R_f\fof{\SI{192}{Hz}}{=}R_2$ between $R_2{=}0.1$ and $R_2{=}100$.
$R_2{<}1$ indicates reduction.
We note again that we refer to energy site amplification. Amplitude site amplification is the square root of energy site amplification. The functional defining $R_f$ is a power law, where the frequency $f$ is either in the base ($R_f{\sim} f^x$) or in the exponent ($R_f{\sim} y^f$). Both relationships are displayed in 
Figure~\ref{fig:testRf}a. 
The models parameterize the linear (purple) and exponential (green) frequency dependence between $\ln R_f$ and $\ln f$, respectively, that is, as said, observed at some sites.
This linear or exponential $R_f$ function is now assumed to represent the reference site characteristics, in contrast to the flat MALM and RUSK response. 
As a result, the $R(f)$ terms obtained for all stations in Figure~\ref{fig:sites} are modulated by this shape. 
Correspondingly, the source spectra shapes $W(f)$ are divided by the $R_f$ test functions.
\par
\begin{figure}
\includegraphics[width=\textwidth]{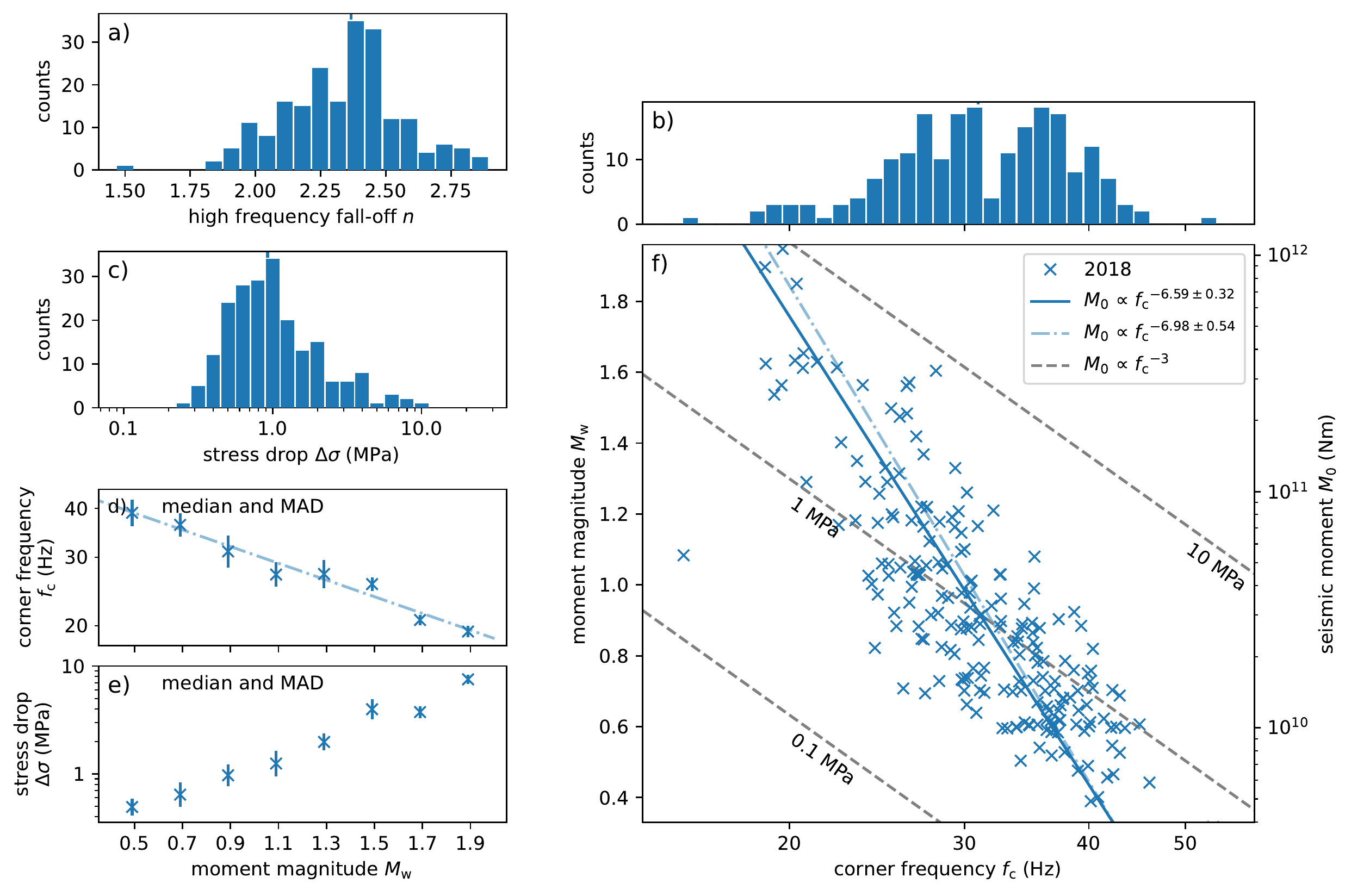}
\caption{Results of the source parameter and source scaling analysis using an exponential reference site model with $R_2{=}10$. The panels correspond to Figure~\ref{fig:fc}. The $M_0$-\fc scaling exponents indicated in the legend in panel (f) correspond to the green indicated slope values at $R_2{=}10$ in Figure~\ref{fig:testRf}d. 
}
\label{fig:testRfexample}
\end{figure}

Figure~\ref{fig:testRf}b shows the trade-off between high-frequency falloff $n$ and $R_2$. 
Again, the $n$ values result from fitting the free source model parameters $n$, \fc, and $M_0$ (Equation~\ref{eq:sourcemodel}) to source spectra shapes $W(f)$ that are obtained from an inversion for which the average reference site term follows the corresponding $R_2$ dependent model.
Here as in the original analysis, the attenuation  is estimated from the coda decay properties and does not affect the $R(f)$-$W(f)$ trade-off.
As an example, Figure~\ref{fig:testRfexample} illustrates the results of this analysis for the exponential model with $R_2{=}10$.
Of course, for both site term models in Figure~\ref{fig:testRf}b, the value for $R_2{=}1$ yields the $n{=}1.74$ value applied in the original analysis.
The high-frequency falloff $n{=}2$ that corresponds to the omega-square model is obtained for $R_2{=}50$ in the linear model, and in the exponential model at $R_2{=}2$.
\par
Figure~\ref{fig:testRf}c displays the trade-off between $R_2$ and the median corner frequency \fc as well as the median stress drop $\Delta\sigma$ for a fix $n{=}2$. The dependence of \fc on $R_2$ shows a negative trend. Similar to 
Figure~\ref{fig:testRf}b the trade-off is larger for the exponential model, i.e., the range of \fc values is larger over the tested range of $R_2$ values. 
A comparison to Figure~\ref{fig:testn}a illustrates the non-linearity of the problem when the site and hence source terms are modulated.
Figure~\ref{fig:testn}a shows that an increase in $n$ leads to an increase in \fc, a steeper slope `pulls' the corner frequency out to larger frequency values.
Taking $n$ as a proxy for $R_2$ as suggested by Figure~\ref{fig:testRf}b, one could thus expect \fc to increase with $R_2$ in Figure~\ref{fig:testRf}c.
The observed opposite decreasing behavior results from the source spectra change, i.e., the simple `pulling' interpretation can not be applied.
\par
Most important for our discussion,
Figure~\ref{fig:testRf}d shows the scaling between corner frequency and seismic moment as a function of $R_2$. For the linear purple model the value of the slope, i.e., of the exponent in the $M_0$-\fc relationship 
(Figures~\ref{fig:fc}, \ref{fig:testRfexample}), increases from around $-7$ at $R_2{=}0.1$ to $-4.6$ for the standard fit and to $-5.4$ for the binned fit at $R_2{=}100$. For the exponential green model the curves show the opposite trade-off---a systematic decrease of the scaling exponent with increasing $R_2$ from approximately $-5$ at $R_2{=}0.1$ to a level around $-9$ at $R_2{=}100$.
To accommodate the fitting of the employed  spectral source model the high-frequency falloff needs to be significantly larger than $n{=}2$ (Figure~\ref{fig:testRf}b).
Thus neither the linear nor the exponential reference site response model makes the inferred $M_0$-\fc scaling relationships convincingly more compatible with self-similarity.
\par
The tests performed by \citet{Trugman2017} and \citet{Trugman2020} using the spectral decomposition method together with data from Southern California and the 2019 Ridgecrest earthquake sequence, respectively,
are similar to our tests presented here. \citet{Trugman2017} fix the high-frequency falloff value and evaluate the trade-off between the frequency dependence of the site amplification and the scaling slope. 
Both \citet{Trugman2017} and \citet{Trugman2020} find the same trade-off as we determined here for the exponential model---slopes that are lower compared to self-similar scaling for larger values of $n$ or $R_2$ as indicated by the trend of the green curve in Figure~\ref{fig:testRf}d.
\par
We consider that a review of all choices, assumptions, and test results together does support the interpretation of our original analysis.
We conclude the available evidence indicates non-self-similar earthquake scaling for the observed magnitude range.

\paragraph{Open research}
\begin{footnotesize}
We use the Qopen code provided at \url{https://github.com/trichter/qopen} \citep{Qopen}.
This research can be reproduced with the Qopen configuration, scripts, and Python source code published at \url{https://github.com/trichter/qopen_finland} \citep{Eulenfeld2023_sourcecode}. Results in electronic format, figures from the main text, and additional figures similar to Figure~\ref{fig:fc} for all tests in the appendices are archived at \citet{Eulenfeld2023_sourcecode}.
Data processing and plotting was performed with the libraries ObsPy, NumPy and matplotlib \citep{Megies2011, NumPy, Hunter2007}.
Waveform data and metadata for the 2018 events can be accessed from \citet{HELdata2018IMS} and \citet{HELdata2018ISUH}.
Waveform data and metadata for the 2020 events can be accessed from \citet{HELdata2020ISUH}.

\end{footnotesize}

\paragraph{Acknowledgments}
\begin{footnotesize}
This work is supported by an Academy of Finland grant, decision number 337913.
The 2018 and 2020 temporary deployments were supported by the Geophysical Instrument Pool Potsdam \citep{HillersData2019,HillersData2021} under the grants 201802 and 201925-ORS2.
We thank editor R. Abercrombie and two anonymous reviewers for comments that helped to improve the manuscript.
G.H. and T.V. appreciate support from staff of the Institute of Seismology and from members of the GeoHel research program of the University of Helsinki.
\end{footnotesize}

\catcode`\^^M=5

\begin{footnotesize}
\setlength{\bibsep}{1.5ex}

\end{footnotesize}

\end{document}